\documentclass[sigconf]{acmart}

\acmConference[Conference acronym 'XX]{Make sure to enter the correct
  conference title from your rights confirmation emai}{June 03--05,
  2018}{Woodstock, NY}
\acmISBN{978-1-4503-XXXX-X/18/06}

\theoremstyle{definition}
\newtheorem{problem}{Problem}

\usepackage{enumerate}
\usepackage{enumitem}

\usepackage{tikz}
\usepackage[normalem]{ulem}
\usetikzlibrary{shapes.geometric, arrows}
\usepackage{bm}

\tikzstyle{process} = [rectangle, rounded corners, minimum width=2cm, minimum height=1cm, text centered, draw=black, fill=white]
\tikzstyle{arrow} = [thick,->,>=stealth]
\tikzstyle{filtering} = [thick,-,>=stealth, dashed, blue]
\tikzstyle{verification} = [thick,-,>=stealth, dashed, green]

\usepackage{multirow}

\begin{document}
\title{Auto-Configuring Entity Resolution Pipelines}

\author{Konstantinos Nikoletos}
\orcid{0000-0003-3465-1197}
\affiliation{%
  \institution{University of Athens}
  \city{Athens}
  \state{Greece}
}
\email{k.nikoletos@di.uoa.gr}

\author{Vasilis Efthymiou}
\orcid{}
\affiliation{%
  \institution{Harokopio University of Athens}
  \city{Athens}
  \country{Greece}
}
\email{vefthym@hua.gr}

\author{George Papadakis}
\orcid{}
\affiliation{%
  \institution{University of Athens}
  \city{Athens}
  \country{Greece}
}
\email{gpapadis@di.uoa.gr}

\author{Kostas Stefanidis}
\affiliation{%
  \institution{Tampere University}
  \city{Tampere}
  \country{Finland}
}
\email{konstantinos.stefanidis@tuni.fi}

\begin{abstract}
The same real-world entity (e.g., a movie, a restaurant, a person) may be described in various ways on different datasets. Entity Resolution (ER) aims to find such different descriptions of the same entity, this way improving data quality and, therefore, data value. However, an ER pipeline typically involves several steps (e.g., blocking, similarity estimation, clustering), with each step requiring its own configurations and tuning. The choice of the best configuration, among a vast number of possible combinations, is a dataset-specific and labor-intensive task both for novice and expert users, while it often requires some ground truth knowledge of real matches. In this work, we 
examine ways of 
automatically configuring 
a state-of-the-art end-to-end ER pipeline based on pre-trained language models under two settings: (i) When ground truth is available. In this case,  
sampling strategies that are typically used for hyperparameter optimization can significantly restrict the search of the configuration space. We experimentally compare their relative effectiveness and time efficiency, applying them to ER pipelines for the first time. (ii) When no ground truth is available. In this case, labelled data extracted from other datasets with available ground truth can be used to train a regression model that predicts the relative effectiveness of parameter configurations. Experimenting with 11 ER benchmark datasets, we evaluate the relative performance of existing techniques that address each problem, but have not been applied to ER before.
\end{abstract}

\maketitle


\ifdefempty{\vldbavailabilityurl}{}{
\vspace{.3cm}
\begingroup\small\noindent\raggedright\textbf{Artifact Availability:}\\
The source code and data
are publicly available at \url{https://github.com/AI-team-UoA/AutoER}.
\endgroup
}

\section{Introduction}\label{sec:introduction}
Entity Resolution (ER) is the problem of identifying different entity descriptions that pertain to the same real-world object (e.g., a person, location, or movie)
\cite{DBLP:journals/csur/ChristophidesEP21}. ER, as a data cleaning method, increases the quality and, subsequently, the value of the data, which can later be used further for downstream applications, like training a machine learning model. To this end, the available data is processed by end-to-end ER pipelines that consist of 
several steps, such as blocking~\cite{DBLP:journals/csur/PapadakisSTP20}, similarity estimation, and clustering~\cite{DBLP:journals/pvldb/HassanzadehCML09,DBLP:journals/vldb/PapadakisETHC23}. Each step
involves multiple configuration parameters, such as 
several blocking strategies, language models, and clustering algorithms to choose from. However, there is no single configuration that dominates the others, as each ER setting (e.g., dataset characteristics, assumptions) has different requirements~\cite{DBLP:journals/csur/ChristophidesEP21,DBLP:journals/pvldb/SunZHWCAL20,DBLP:journals/pvldb/KopckeTR10}. 

\begin{figure*}[t]
    \centering
    \includegraphics[width=0.95\linewidth]{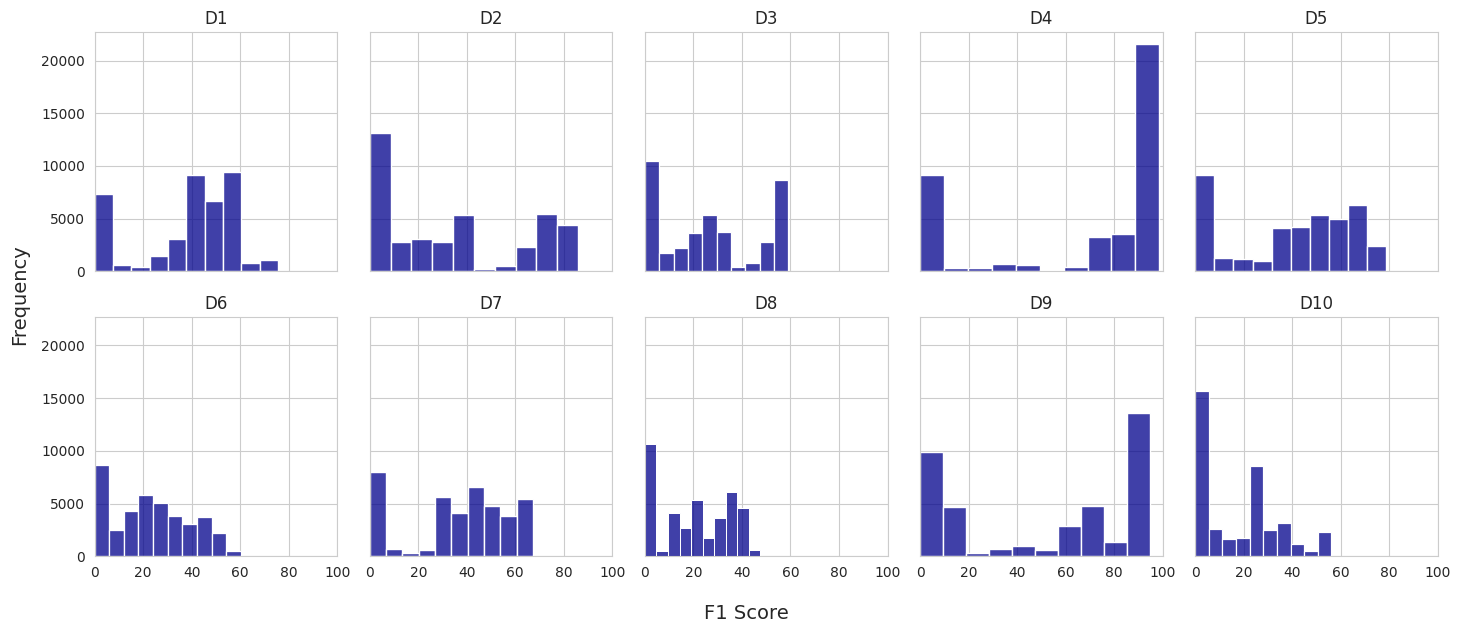}
    \vspace{-10pt}
    \caption{The distribution of F1-scores for the end-to-end ER pipeline of Figure~\ref{fig:eeter_pipeline}, when applying the 39,900 different configurations of Table~\ref{tab:parameter-values} to the 10 real-world datasets of Table~\ref{tab:dataset-specs}.
    }
    \label{fig:f1_boxplot_all}
    \vspace{-14pt}
\end{figure*}

This is illustrated in Figure \ref{fig:f1_boxplot_all}, which demonstrates the range of F1 scores over 10 established real-world datasets 
(see Section \ref{sec:expSetup} for more details). For each dataset, we assessed the performance of the state-of-the-art end-to-end ER pipeline in Figure \ref{fig:eeter_pipeline} \cite{DBLP:journals/pvldb/ZeakisPSK23} through a grid search over all configuration parameters in Table~\ref{tab:parameter-values} (see Section \ref{sec:eteer_pipeline} for more details). In total, we applied 39,900 different configurations per dataset, with their performance exhibiting a large deviation between the maximum and the minimum F1 score: in most datasets, the vast majority of configurations covers all values in $[0,0.5]$, while in datasets $D_4$ and $D_9$, the range is even larger, going up to 1.0 (both datasets contain unambiguous bibliographic data like author names and publication titles that ensure high performance for a large portion of configurations). This indicates significant sensitivity to the configuration parameters of the ER pipeline, with most options typically leading to rather low effectiveness. 

These settings highlight the importance of fine-tuning end-to-end ER pipelines.
However, choosing the right configuration is a non-trivial task that requires time-consuming trial-and-error experimentations even for experts. To address this problem, we formulate parameter fine-tuning as a regression task over a large search space and leverage established sampling-based techniques for hyperparameter optimization. We experimentally demonstrate that these techniques are able to propose parameter configurations that approximate the optimal performance by curtailing the search space by orders of magnitude -- with fewer than 100 trials.

Another obstacle in practical ER applications is the lack of ground truth with correct matches that allows for
evaluating the performance of different configurations automatically.
To address this task, we formulate a regression task, where the input comprises a feature vector describing the combination of a dataset and an ER pipeline configuration, while the output is the corresponding F1 score. In this context, datasets with known ground truth act as training instances, while datasets without a ground truth form the testing instances. We apply the trained model to all instances of a dataset without a ground truth and the one maximizing the expected F1 score indicates the most suitable parameter configuration. Using the end-to-end ER pipeline of Figure \ref{fig:eeter_pipeline}, we apply existing techniques for each part of this regression model: feature engineering is based on dataset profiling, the instances are generated by grid search, sampling-based algorithms, or both, and the learning process is carried out by two established approaches with integrated feature selection, namely Random Forests and AutoML. Our extensive experimental evaluation demonstrates which combination of these techniques yields the best performance.

In summary, the contributions of this work are the following: 
\begin{itemize}[leftmargin=*]
    \item We define two problems for the automatic configuration of end-to-end ER pipelines: one assuming a ground truth and another one assuming no ground truth information.
    \item We show how both problems can be addressed by existing techniques that have never been applied to the task of automatic ER pipeline configuration.
    \item We perform extensive experimental evaluations over 11 established, real-world datasets 
    to identify the best solutions per problem.
\end{itemize}

We have open-sourced our code with a permissive license on Github: \url{https://github.com/AI-team-UoA/AutoER}.


\section{Problem Definition}\label{sec:problem}


Let an \emph{entity collection} $\mathcal{E}$ be a set of entity descriptions, where each description (also called ``entity'' for brevity)
is a set of attribute-value pairs representing a real-world entity (e.g., an entity description can be a database record, a row in a csv file, a json element or an ontology class instance). 

End-to-end ER (ETEER) is the problem of receiving one or more entity collections and returning a set of entity \emph{clusters}, with each cluster corresponding to a set of entity descriptions that refer to the same real-world entity. 
More formally:

\begin{definition}[End-to-end Entity Resolution (ETEER)]\label{def:eteer}
Given two entity collections $\mathcal{E}_1$ and $\mathcal{E}_2$, return a set of disjoint clusters $\mathcal{C}$, such that 
$\bigcup_{c_l \in \mathcal{C}}c_l = \mathcal{E}_1 \cup \mathcal{E}_2$, and
$\forall c_l \in \mathcal{C}, 
 e_i \in \mathcal{E}_1 \cap c_l, e_j \in \mathcal{E}_2 \cap c_l \Rightarrow e_i \equiv e_j$, 
where $e_i \equiv e_j$ denotes that the two entity descriptions~\emph{match}.
\end{definition}


This definition applies to Record Linkage \cite{DBLP:books/daglib/0030287}, also known as Clean-Clean ER~\cite{DBLP:journals/vldb/PapadakisETHC23}, where $\mathcal{E}_1$ and $\mathcal{E}_2$ are individually duplicate-free, but overlapping, sharing some entities. In this setting, each cluster contains either a single  entity from one entity collection, or 
one from each collection. Definition~\ref{def:eteer} can be easily generalized to Dirty ER \cite{DBLP:journals/pvldb/HassanzadehCML09}, also known as Deduplication \cite{DBLP:books/daglib/0030287}, where the input comprises a single entity collection with duplicates in itself. In this case, the resulting clusters are also disjoint, but there is no limit on their size, i.e., the number of entities they contain. It can also be generalized to the case of Multi-Source ER \cite{DBLP:conf/esws/SaeediPR20,DBLP:conf/adbis/SaeediPR17}, where the input comprises multiple, duplicate-free entity collections. This setting requires that each cluster cannot contain more than one entity from the same collection. 


The problem of Definition~\ref{def:eteer} is typically solved through an \emph{ETEER pipeline}, also called \emph{workflow}, denoted by $w(\mathcal{E}_1, \mathcal{E}_2)$. This involves a series of processing steps, which potentially require the configuration of some parameters. More formally:

\begin{definition}[ETEER Pipeline]
An ETEER Pipeline $w(\mathcal{E}_1, \mathcal{E}_2) = (S, P, V)$ is a series of processing steps $S = s_1, s_2, \ldots, s_n$ for solving ETEER, with each processing step $s_i$ potentially requiring the configuration of some parameters $P_i = p_1^i, p_2^i, \ldots, p_m^i$, and with each parameter $p_j^i$ accepting a set of possible values $V_j^i$.
\end{definition}

Following the previous definition, we 
call the set $P = \bigcup\limits_{i \in [1..n]}
P_i$ \emph{configuration parameters} and 
the set $V = \bigcup\limits_{P_i \in P, p_j \in P_i}V_j^i$ as \emph{configuration values} or \emph{possible values}.

The \underline{effectiveness} of an ETEER pipeline is assessed in terms of precision, recall, and F1-score, with respect to the ground truth of known matches $\mathcal{D}$. In more detail,
\textit{precision} is the portion of clusters in $\mathcal{C}$ which contain a pair of entities that also appears in $\mathcal{D}$; 
\textit{recall} is the portion of pairs in $\mathcal{D}$ that co-occur in some cluster of $\mathcal{C}$;
\textit{F1-score} is the harmonic mean of precision and recall.

The \underline{time efficiency} of an ETEER pipeline can be measured in terms of the 
overall runtime, i.e., the time between receiving $\mathcal{E}_1$ and $\mathcal{E}_2$ and producing $\mathcal{C}$. 
In this work, we also consider
the \emph{search time}, i.e., the time needed to 
recommend configuration values.

In this context, we address two problems for fine-tuning ETEER pipelines, based on whether a ground truth is available (Problem~\ref{pr:pr1}) or not (Problem~\ref{pr:pr2}).
For the evaluation, we always use the ground truth of known matches $\mathcal{D}$. In Problem~\ref{pr:pr1}, we also use $\mathcal{D}$
for tuning the configuration parameters, unlike Problem~\ref{pr:pr2}.
More formally:


\begin{figure}[t]
    \centering
    \includegraphics[width=\linewidth]{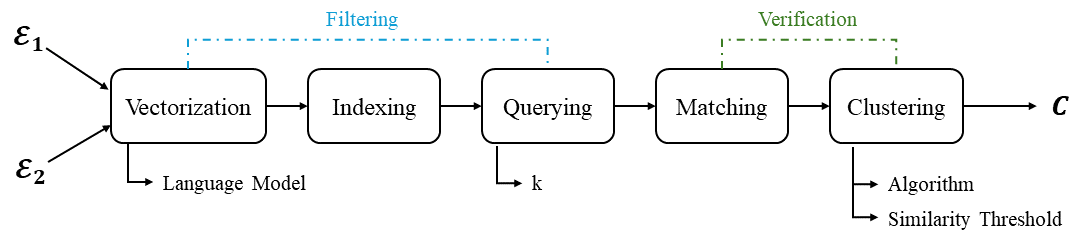}
    \vspace{-20pt}
    \caption{The ETEER pipeline considered in this work.}
    \label{fig:eeter_pipeline}
    \vspace{-12pt}
\end{figure}

\begin{problem}
\label{pr:pr1}
Given an ETEER pipeline $w(\mathcal{E}_1, \mathcal{E}_2)$ for two entity collections $\mathcal{E}_1$ and $\mathcal{E}_2$, along with a subset of the ground truth of matches $\mathcal{D}' \subseteq \mathcal{D}$, return the configuration values $V' \subseteq V$ of $w(\mathcal{E}_1, \mathcal{E}_2)$ that maximize the effectiveness of $w(\mathcal{E}_1, \mathcal{E}_2)$ in terms of F1-score, while minimizing the search time.
\end{problem}
For the returned configuration values $V'$, it holds that $|V'| = |P|$, i.e., one value should be returned for each configuration parameter.

The na\"ive approach is to apply \textit{grid search}, considering all meaningful values for all configuration parameters in the specified pipeline. However, this is impractical and time-consuming, due to the enormous configuration space.
More advanced algorithms for configuration optimization are required to effectively restrict the search to a small portion of the configuration space.


A more difficult variation of the first problem is to fine-tune a specific ETEER pipeline without having any portion of the real matches $\mathcal{D}$ provided as input. This task can be formalized as follows:

\begin{problem}
\label{pr:pr2}
Given an ETEER pipeline $w(\mathcal{E}_1, \mathcal{E}_2)$ for two entity collections $\mathcal{E}_1$ and $\mathcal{E}_2$, return the configuration values $V' \subseteq V$ of $w(\mathcal{E}_1, \mathcal{E}_2)$ that maximize the effectiveness of $w(\mathcal{E}_1, \mathcal{E}_2)$ in terms of F1-score, while minimizing the search time.
\end{problem}

To the best of our knowledge, no prior work on ER examines these tasks, even though both can be solved by existing techniques. 
\section{ETEER pipeline}\label{sec:eteer_pipeline}

We now present the ETEER pipeline that we use to address
Problems~\ref{pr:pr1} and \ref{pr:pr2}. 
Due to the quadratic complexity of the brute-force approach, it consists of two steps:
\begin{itemize}[leftmargin=*]
    \item \emph{Filtering}, which reduces the computational cost by identifying a set of promising \emph{candidate pairs} to be matched.
    \item \emph{Verification}, which analytically examines each pair of candidates.
\end{itemize}

Based on a recent experimental study \cite{DBLP:journals/pvldb/ZeakisPSK23}, our solution relies on the ETEER pipeline shown in Figure~\ref{fig:eeter_pipeline}, which leverages language models. This approach not only combines high effectiveness with high time efficiency, but also requires the fine-tuning of a limited set of configuration parameters.




The first step of this ETEER pipeline is Vectorization, where the input entities are transformed into embedding vectors using pre-trained LMs, either static ones like Word2Vec \cite{DBLP:journals/corr/abs-1301-3781} and GloVe \cite{DBLP:conf/emnlp/PenningtonSM14}, or dynamic ones, like BERT \cite{DBLP:conf/naacl/DevlinCLT19} and SentenceBERT~\cite{DBLP:conf/emnlp/ReimersG19}.
The former are context-agnostic, assigning each word to the same vector regardless of its context, while the latter are context-aware, generating different embeddings for different meanings of the same word (e.g., trunk as part of a tree or an elephant). Therefore, selecting the appropriate language model is crucial for achieving good results. Following \cite{DBLP:journals/pvldb/ZeakisPSK23}, we restrict our configurations to five state-of-the-art SentenceBERT models and two static models, listed in Table \ref{tab:parameter-values} 
\textbf{(Parameter: LM)}. 
Pre-trained BERT models yield scores of very low distinctiveness, failing to distinguish between matching and non-matching pairs, unless
they are fine-tuned \cite{DBLP:journals/pvldb/ZeakisPSK23}.
Describing them in more detail goes beyond the scope of this work.

\begin{table}[t]
    \centering
    \caption{Configuration parameters and their possible values.}
    \vspace{-8pt}
    \label{tab:parameter-values}
    \begin{tabular}{|p{3cm}|p{4.5cm}|}
        \hline
         \multicolumn{1}{|c|}{\textbf{Parameters ($P$)}} & \multicolumn{1}{|c|}{\textbf{Possible values ($V$)}} \\ \hline
        Language model (LM) & smpnet, S-GTR-T5, sdistilroberta, sminilm, sent\_glove, fasttext, word2vec \\ \hline
        $k$ & $[1, 100]$ with a step of 1\\ \hline
        Clustering algorithm & Unique Mapping Clustering (UMC), Kiraly Clustering (KC), Connected Components (CCC) \\ \hline
        Similarity Threshold & $[0.05, 0.95]$ with a step of 0.05\\ \hline
    \end{tabular}
    \vspace{-14pt}
\end{table}

Note that we apply a schema-agnostic serialization scheme, which simply concatenates all attribute values into a sentence describing each entity (i.e., all attribute names are excluded). This way, we inherently address noise in the form of misplaced values, while also achieving very high performance under versatile settings~\cite{DBLP:journals/pvldb/ZeakisPSK23}.

The second step is Indexing. It receives as input all embedding vectors of $\mathcal{E}_1$ and organizes them in a data structure that facilitates their retrieval in descending distance from a given query, i.e., an embedding vector of $\mathcal{E}_2$. 
Based on a recent experimental study \cite{DBLP:journals/is/AumullerBF20}, we employ 
FAISS \cite{DBLP:journals/corr/abs-2401-08281} for this purpose, due to its excellent balance between effectiveness and time efficiency.
We actually use a FAISS configuration that returns exact results, which does not require parameter tuning. 
For datasets with more than 10,000 entities, the approximate search should be applied to ensure high time efficiency, but its configuration is also straightforward.\footnote{For more details, please refer to \url{https://github.com/facebookresearch/faiss/wiki/Guidelines-to-choose-an-index}.}

The third step is Querying. It receives as input all embedding vectors of $\mathcal{E}_2$ and uses each vector/entity as a query to the FAISS index. The result of each query consists of the $k$ most similar entities from $\mathcal{E}_1$ in terms of cosine similarity. 
The $k$ parameter is the sole configuration parameter of this step, playing a major role in the performance of ETEER. The higher the value of $k$, the higher the filtering recall, at the cost of lower precision. 
Note that the filtering recall sets the upper bound of the overall ETEER recall, given that the subsequent steps do not detect new matches.
Hence, $k$ is the second configuration parameter of the pipeline \textbf{(Parameter:~$\bm{k}$)}.

The first three steps together implement the Filtering approach, which reduces the computational cost to the $k$ most similar $\mathcal{E}_2$ entities per $\mathcal{E}_1$ entity. 
It receives as input the given entity collections and returns as output a set of candidate pairs.


Subsequently, the set of candidate pairs 
is transformed into a \textit{similarity graph}, where every node is an entity and every edge represents a candidate pair, weighted according to the cosine similarity returned by FAISS. The similarity graph is bipartite in the Record Linkage settings we are considering. This transformation is carried out by Matching, the fourth step of our ETEER pipeline. No configuration parameter is involved in this process. 

The final step of the ETEER pipeline is Clustering, which applies bipartite graph matching algorithms to the similarity graph generated by the previous step. Based on a recent experimental study~\cite{DBLP:journals/vldb/PapadakisETHC23}, three established algorithms combine high time efficiency with high effectiveness: Connected Components (i.e., transitive closure), Unique Mapping Clustering \cite{DBLP:conf/kdd/Lacoste-JulienPDKGG13}, and Kiraly Clustering~\cite{DBLP:journals/algorithms/Kiraly13} \textbf{(Parameter: Clustering)}.\footnote{A more detailed description of their functionality lies out of the scope of this work, but the interested reader can refer to \cite{DBLP:journals/vldb/PapadakisETHC23} and the original publications of each algorithm.}
All algorithms are configured by 
a similarity threshold \textbf{(Parameter: Threshold)}, which prunes the edges with a lower weight.

Note that Clustering is the only step that 
depends on the type of ER task at hand. In the case of Dirty ER, the state-of-the-art unconstrained algorithms for undirected similarity graphs are experimentally evaluated in \cite{DBLP:journals/pvldb/HassanzadehCML09}. Among them, Markov Clustering exhibits a performance that is robust to noise and portion of duplicates, while achieving high effectiveness and scalability. For Multi-source ER, the main clustering techniques are evaluated in \cite{DBLP:journals/csimq/SaeediNPR18}, with CLIP~\cite{DBLP:conf/esws/SaeediPR18} outperforming all others -- CLIP essentially generalizes Unique Mapping Clustering to Multi-source~ER.

The configuration space of our pipeline is summarized in Table~\ref{tab:parameter-values}. Note that grid search should consider 39,900 different parameter combinations per dataset in order to maximize the corresponding F1 (7 pre-trained language models $\times$ 100 values for $k$ $\times$ 3 clustering algorithms $\times$ 19 values for similarity threshold). Our goal is to experimentally evaluate algorithms that converge to the maximum performance, while minimizing the trials in this search space.
\section{Tackling Problem \ref{pr:pr1}}
\label{sec:problem-1}

Fine-tuning an ETEER pipeline based on the ground truth comprising all duplicate pairs is similar to hyperparameter fine-tuning in Machine and Deep Learning models \cite{DBLP:journals/ijon/YangS20}. The only difference is the form of the data: in the latter case, the data comes in the form of positive and negative labelled instances, which is not possible in ETEER, due to the extremely large number of pairs that arise even from relatively small datasets with few thousand entities in $\mathcal{E}_1$ and $\mathcal{E}_2$. Besides, the number of positive pairs grows linearly with respect to the number of given entities, while all other pairs are negative, with their number growing quadratically \cite{DBLP:journals/pvldb/GetoorM12}. Therefore, instead of enumerating and labelling all possible pairs in the Cartesian product of $\mathcal{E}_1 \times \mathcal{E}_2$, hyperparameter fine-tuning is guided by the F1-score corresponding 
to each parameter configuration of the ETEER pipeline to be fine-tuned. This is the only change that needs to be applied to existing algorithms for hyperparameter fine-tuning, yet they have not been applied before to Problem 1.

Typically, hyperparameter fine-tuning is carried out by \emph{sampling techniques}, which aim at optimizing the balance between effectiveness and time efficiency:  
their goal is to maximize efficiency by reducing the search space to the most promising subset of configuration parameters, without leaving out the ones maximizing effectiveness~\cite{OPTUNA}. 
These sampling techniques are categorized into two main types: 
the \textit{relational} ones select the next set of configuration parameters based on the correlations between them, while the \textit{independent} ones disregard inter-parameter correlations.

In this work, we consider the following state-of-the-art sampling algorithms \cite{OPTUNA}, which cover both types of sampling techniques:

(1)  \emph{RandomSampler} \cite{QMCSampler}. As its name suggests,  
    hyper-parameter values are chosen randomly from the specified search space. This is an independent method that does not use any prior knowledge or adaptive mechanisms to guide the search process. Instead, it 
    explores the search space in a purely stochastic manner, i.e.,
    each point in the search space has an equal probability of being explored. This approach results in a uniform coverage of the search space. Even in high-dimensional spaces, random sampling is capable of identifying near-optimal configurations with fewer trials, especially when compared to the grid search approach. 

(2) Tree-structured Parzen Estimator (\emph{TPESampler}) \cite{TPE1, TPE2, TPE3}.
    During each trial for every parameter, this independent algorithm fits two Gaussian Mixture Models (GMMs): \( l(x) \) to the set of parameter values with the best objective outcomes, and \( g(x) \) to the remaining parameter values. It then selects the parameter value \( x \) that maximizes the ratio \( l(x) / g(x) \).  In other words, TPESampler models the distribution of good and bad hyperparameters using non-parametric density estimators, focusing on promising regions of the search space. By iteratively updating these models based on observed results, TPESampler may need more trials to explore potentially optimal areas, but typically exhibits better search performance than grid search and RandomSampler. 

(3) Quasi Monte Carlo Sampler (\emph{QMCSampler}) \cite{QMCSampler}. This independent approach uses quasi-random methods, specifically low-discrepancy sequences, to enhance the efficiency of hyperparameter optimization by ensuring uniform coverage of the search space. Unlike RandomSampler, which can have uneven distribution, low-discrepancy sequences spread points more evenly, avoiding clumps and gaps. This uniformity increases the probability of finding optimal hyperparameters with fewer trials, particularly in high-dimensional settings where important dimensions need adequate sampling. 

(4) Gaussian Process-based Bayesian Sampler (\emph{GPSampler}) \cite{NIPS2012_GPSampler, QMCSampler}. This is a relational sampler that fits a Gaussian process to the objective function and optimizes the acquisition function to suggest the next set of parameters. First,
    it constructs a probabilistic model of the objective function using a Gaussian process and then it exploits this model to estimate the most promising regions in the search space. 
    To this end, it employs the log expected improvement 
    in combination with QMCSampler
    to optimize the acquisition function (note that in Bayesian optimization, the acquisition function tackles the exploration-exploitation trade-off, efficiently providing numerical estimations that indicate the most promising configuration parameters to be tested \cite{DBLP:conf/nips/WilsonHD18}).
    Typically, GPSampler is highly effective in hyperparameter optimization,
    as it leverages the dependencies between hyperparameters and avoids redundant trials, due to their probabilistic estimation of the objective function \cite{NIPS2012_GPSampler, QMCSampler}. 

Regarding the time complexity \textit{per trial} of these samplers, we use $d$ for the dimensionality of the search space and $n$ for the number of completed trials. The most efficient algorithm is RandomSampler, with $O(d)$, followed by QMCSampler with $O(d n )$, TPESampler with $O(d n logn)$ and GPSampler with $O(n^3)$ -- for more details, please refer to {\footnotesize \url{https://optuna.readthedocs.io/en/stable/reference/samplers/index.html}}.

To the best of our knowledge, none of these algorithms has been applied to fine-tuning ETEER pipelines, addressing Problem 1. More specifically, each algorithm is applied to Problem 1 as follows: 
\begin{enumerate}[leftmargin=*]
    \item It receives $V$ as input, i.e., the domain of each configuration parameter, along with the maximum number of trials.
    \item In each trial, it selects a value for each configuration parameter.
    \item The resulting ETEER pipeline is applied to the given dataset to estimate its performance with respect to F1-score, using the available ground truth.
    \item After consuming the budget of trials, the configuration values $V' \subseteq V$ maximizing F1-score are returned as output.
\end{enumerate}

\begin{figure*}[h!]
    \centering
    \includegraphics[width=0.97\textwidth]{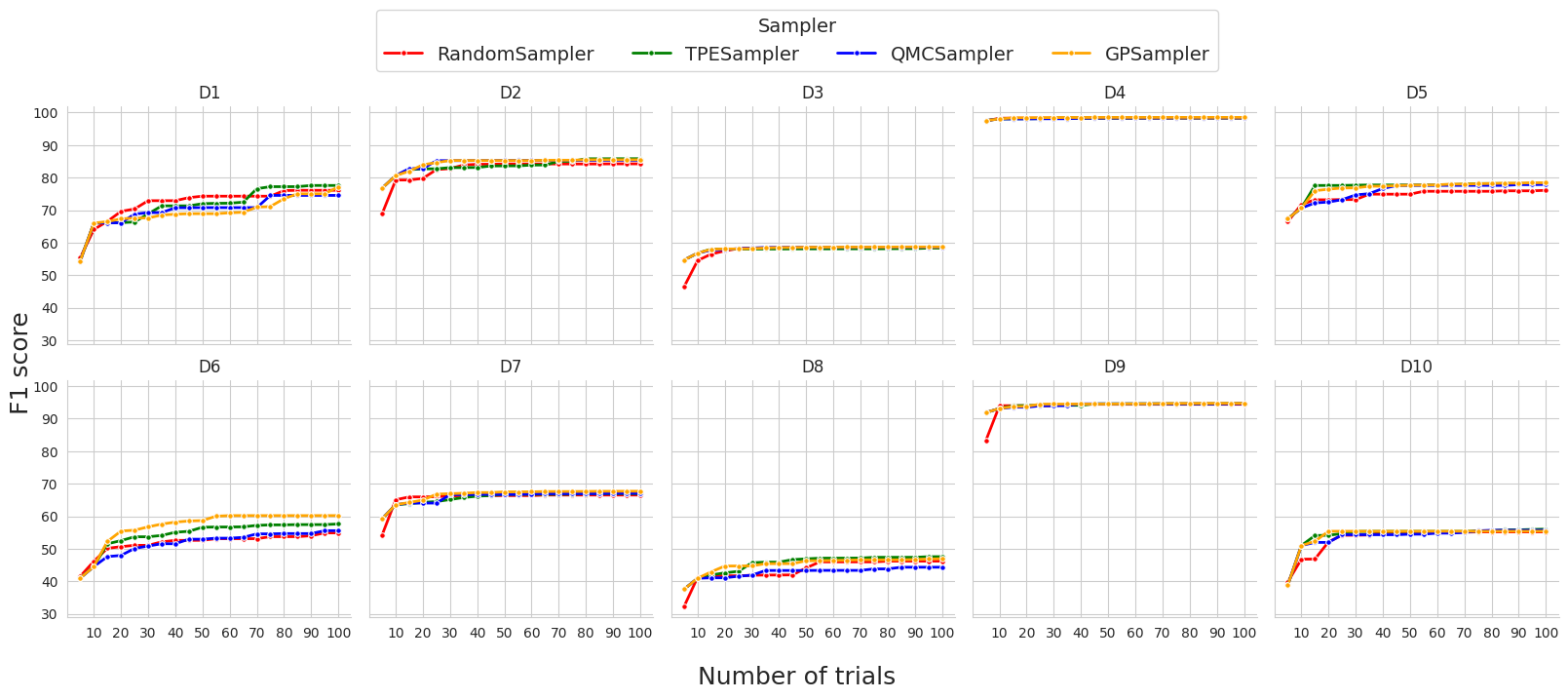}
    \vspace{-10pt}
    \caption{F1-score of the sampling-based search algorithms in Section \ref{sec:problem-1} as the number of trials increases from 5 to 100.}
    \label{fig:samplers-convergence-avg-f1}
    \vspace{-10pt}
\end{figure*}

\section{Experimental Evaluation - Problem 1}\label{sec:experiments-p1}


\subsection{Experimental setup}
\label{sec:expSetup}
All experiments were implemented in Python, v. 3.9. For the implementation of the ETEER pipeline, we used pyJedAI v. 0.1.8\footnote{\url{https://github.com/AI-team-UoA/pyJedAI}} \cite{Nikoletos2022pyJedAIAL}. 
For the implementation of the sampling algorithms, we used Optuna v. 3.6.1\footnote{\url{https://optuna.org}}. 
All 
experiments 
were executed on a server running 
Ubuntu 22.04, 
with Intel Core i7-9700K @4,9GHz and 68 GB RAM.


\textbf{Datasets.} For Record Linkage, we use 11 publicly available, real-world datasets that are popular in the literature~\cite{DBLP:journals/pvldb/Thirumuruganathan21,DBLP:journals/pvldb/KopckeTR10,DBLP:journals/pvldb/0001SGP16,DBLP:conf/sigmod/MudgalLRDPKDAR18}. Their technical characteristics are reported in Table~\ref{tab:dataset-specs}. $D_{1}$, which was first used in OAEI 2010,
contains restaurant descriptions. $D_{2}$ encompasses duplicate products from the online retailers Abt.com and Buy.com \cite{DBLP:journals/pvldb/KopckeTR10}. $D_{3}$ matches product descriptions from Amazon.com and the Google Base data API (GB) \cite{DBLP:journals/pvldb/KopckeTR10}. $D_{4}$ entails bibliographic data from DBLP and ACM \cite{DBLP:journals/pvldb/KopckeTR10}. $D_{5}$, $D_{6}$ and $D_{7}$ involve descriptions of television shows from TheTVDB.com (TVDB) and of movies from IMDb and themoviedb.org (TMDb) \cite{DBLP:conf/esws/ObraczkaSR21}. $D_{8}$ matches product descriptions from Walmart and Amazon \cite{DBLP:conf/sigmod/MudgalLRDPKDAR18}. $D_{9}$ involves bibliographic data from publications in DBLP and Google Scholar (GS) \cite{DBLP:journals/pvldb/KopckeTR10}. Finally, 
$D_{10}$ interlinks movie descriptions from IMDb and DBpedia \cite{DBLP:journals/is/PapadakisMGSTGB20}, including a different snapshot of IMDb than $D_5$~and~$D_6$. 




\begin{table}[t]
\centering 
\caption{Technical characteristics of the datasets used in our experimental analysis. $|E_x|$ stands for the number of entities in data source $x$ and $|D|$ for the number of duplicates.
} 
\vspace{-10pt}
\label{tab:dataset-specs}
\begin{tabular}{|p{0.9cm}|p{2.1cm}|r|r|r|}
\hline
\multicolumn{1}{|c|}{\textbf{Dataset}} & \multicolumn{1}{|c|}{\textbf{Names}} & \multicolumn{1}{|c|}{\textbf{$\mathbf{|E_1|}$}} & \multicolumn{1}{|c|}{{$\mathbf{|E_2|}$}} & \multicolumn{1}{|c|}{\textbf{$\mathbf{|D|}$}} \\
\hline
\hline
\multirow{2}{*}{D1} & Restaurants1-& \multirow{2}{*}{340} & \multirow{2}{*}{2,257} & \multirow{2}{*}{89} \\
& Restaurants2 & & & \\
\hline
D2 & Abt-Buy & 1077 & 1,076 & 1,076 \\
\hline
\multirow{2}{*}{D3} & Amazon- & \multirow{2}{*}{1,355} & \multirow{2}{*}{3,040} & \multirow{2}{*}{1,103} \\
& Google Products & & & \\
\hline
D4 & DBLP-ACM & 2,617 & 2,295 & 2,225 \\
\hline
D5 & IMDB-TMDB & 5,119 & 6,057 & 1,969 \\
\hline
D6 & IMDB-TVDB & 5,119 & 7,811 & 1,073 \\
\hline
D7 & TMDB-TVDB & 6,057 & 7,811 & 1,096 \\
\hline
\multirow{2}{*}{D8} & Walmart- & \multirow{2}{*}{2,555} & \multirow{2}{*}{22,075} & \multirow{2}{*}{853} \\
& Amazon & & & \\
\hline
D9 & DBLP-Google Scholar & 2,517 & 61,354 & 2,309 \\
\hline
D10 & IMDB-DBpedia & 27,616 & 23,183 & 22,864 \\
\hline
\hline
D11 & DBpedia & 1,190,734 & 2,164,041 & 892,579 \\
\hline
\end{tabular}
\vspace{-10pt}
\end{table}

\begin{table*}[t]
\begin{center}  
\caption{Performance of the two baseline methods for Problem 1: (a) the default workflow (st5, 10, UniqueMappingClustering, 0.5), and (b) the best grid-search trial. For the latter, we also report the parameter configuration.}
\vspace{-8pt}
\label{tab:best-gridesearch-trials}
\begin{tabular}{|c||c|c||c|c|c|c||c|c|c|}
\hline
\multirow{2}{*}{Dataset} & \multicolumn{2}{c||}{Default configuration} & \multicolumn{7}{c|}{Best grid-search configuration} \\
& F1-score & Runtime (sec) & LM & K & Clustering & Threshold & F1-score & Runtime (sec) & Grid search time (hrs) \\
\hline
\hline
D1 & 47.44 & 1.90 & smpnet & 3 & CCC & 0.90 & 75.53 & 2.87 & 41 \\
D2 & 85.85 & 3.28 & st5 & 10 & UMC & 0.35 & 85.85 & 2.51 & 38 \\
D3 & 57.35 & 1.49 & sminilm & 7 & UMC & 0.45 & 59.19 & 1.43 & 45 \\
D4 & 97.56 & 4.45 & st5 & 1 & UMC & 0.80 & 98.60 & 1.43 & 45 \\
D5 & 57.74 & 5.05 & st5 & 6 & KC & 0.75 & 78.73 & 2.98 & 60 \\
D6 & 30.39 & 1.81 & sminilm & 1 & UMC & 0.55 & 60.25 & 0.91 & 71 \\
D7 & 35.68 & 2.21 & sminilm & 93 & CCC & 0.80 & 67.36 & 28.64 & 74 \\
D8 & 35.82 & 4.46 & st5 & 4 & KC & 0.90 & 47.56 & 3.19 & 145 \\
D9 & 92.04 & 13.11 & st5 & 42 & KC & 0.80 & 94.89 & 21.37 & 329 \\
D10 & 53.63 & 27.96 & st5 & 2 & KC & 0.65 & 56.12 & 28.77 & 320 \\
\hline
\end{tabular}
\vspace{-8pt}
\end{center}  
\end{table*}

\subsection{Evaluation Results}
\label{sec:tackleProblem1}
We address three research questions while tackling Problem~1:
\begin{enumerate}[leftmargin=*, label=RQ\arabic*), start=1]
    \item Which of the sampling-based search algorithms 
    in Section \ref{sec:problem-1} converges faster 
    to its maximum performance?
    \item Does sampling-based search outperform the baselines with respect to effectiveness?
    \item Does sampling-based search outperform the baselines with respect to time efficiency?
\end{enumerate}

We use two approaches as baselines: 
\begin{enumerate}[leftmargin=*]
    \item A default configuration based on \cite{DBLP:journals/pvldb/ZeakisPSK23} that combines the s-t5 language model with $k$=10, Unique Mapping Clustering and similarity threshold = 0.5. 
    \item The best performance estimated by grid search among the configuration parameters in Table \ref{tab:parameter-values}. 
\end{enumerate}

The performance of those baselines per dataset with respect to effectiveness and time efficiency is reported in Table \ref{tab:best-gridesearch-trials}.

\textbf{RQ1.}
We evaluate all samplers discussed in Section~\ref{sec:problem-1}, configuring each one 
to run 100 trials per dataset, according to the recommendations from the Optuna documentation\footnote{\url{https://optuna.readthedocs.io/en/stable/reference/samplers/index.html}}. To analyze the convergence behavior of each sampler, we vary the maximum number of trials from 5 to 100, in increments of 5. For each budget of trials, we perform experiments with five different random seeds and consider the average, thus providing a robust estimation of the actual performance per number of trials. The results appear in Figure \ref{fig:samplers-convergence-avg-f1}, with the horizontal axes corresponding to the number of trials and the vertical ones to the respective average F1-score.


For all samplers, convergence is generally observed in around 30 trials for most datasets. This means that with just 30 trials, they approximate the best performance in almost all datasets.
The only exceptions are $D_1$, $D_6$ and $D_8$. As observed in Figure \ref{fig:f1_boxplot_all}, in these datasets the portion of configuration parameters that maximize F1 is rather small. As a result, we need to increase the number of trials to much more than 30 in order to identify a configuration matching or approximating the maximum performance.

Regarding the relative performance of the considered samplers, Figure \ref{fig:samplers-convergence-avg-f1} indicates that there are no significant differences among them. Their behavior is quite similar, determined largely by the dataset at hand. Nevertheless, \textit{GPSampler} shows marginally better results in several cases.

\textbf{RQ2.} We now compare the effectiveness of the four samplers with that of the two baselines in Table \ref{tab:best-gridesearch-trials}. 
For the grid-search baseline, we also report the best-performing configuration.
As expected, the default configuration consistently underperforms grid search to a significant extent. The only exception appears in $D_2$, where they yield the same F1-score, because the grid search configuration is almost identical with the default one (they differ only in the similarity threshold). Hence, for brevity, we exclusively compare the samplers with the best grid search configuration.





\begin{figure*}[t]
    \centering
    \includegraphics[width=0.94\textwidth]{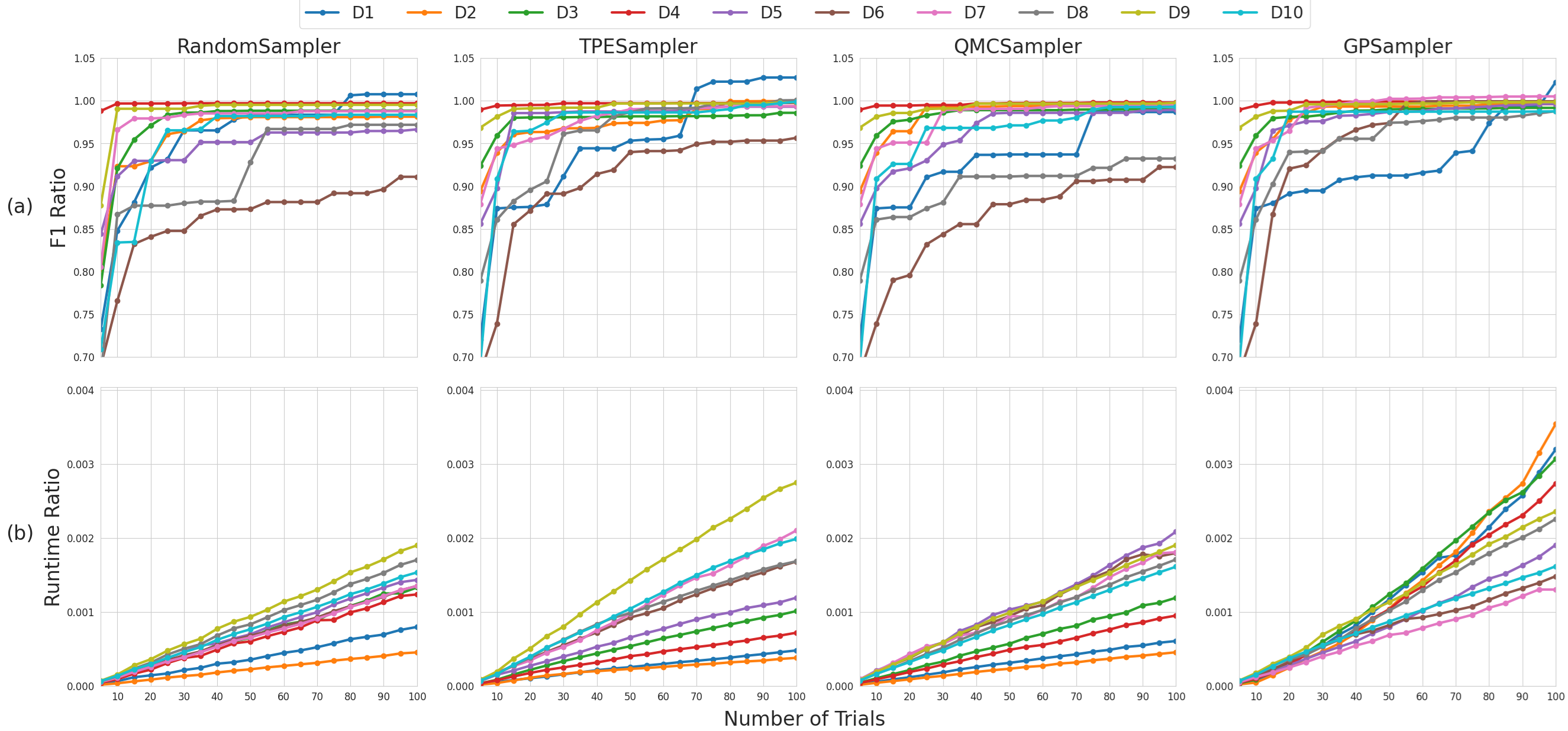}
    \vspace{-10pt}
    \caption{The ratio between (a) the F1-score of sampling-based search algorithms and grid search, and (b) their run-times.
    }
    \vspace{-10pt}
    \label{fig:p1-samplers-performance-f1}
\end{figure*}

This comparison is depicted in Figure \ref{fig:p1-samplers-performance-f1}(a), which contains a separate diagram for each sampler with its convergence to the best grid search performance in Table \ref{tab:best-gridesearch-trials}. The vertical axes 
correspond to \textit{F1 ratio}, which is defined as ``samplerF1''/``gridSearchF1''. That is, an F1 ratio of 1.0 indicates that the two approaches yield the same F-Measure despite their different configuration. Values lower than 1.0 indicate lower sampling-based performance than grid search, while ratios $>1.0$ denote that the sampler identified a configuration that outperforms all those tested by grid search. This should be attributed to the discrete similarity thresholds considered by grid search, unlike samplers, which may consider any value in $[0,1]$.

RandomSampler underperforms grid search in almost all cases. Yet, with just 20 trials, its F1 ratio exceeds 0.90 in all datasets but
$D_1$, $D_6$ and $D_8$.
These three datasets 
convey very few top-performing configurations, as shown in Figure \ref{fig:f1_boxplot_all}. 
Note also that with just 30 trials, its F1 ratio exceeds 0.95 in all datasets but $D_1$, $D_6$ and $D_8$. In $D_1$, RandomSampler converges to grid search after 70 trials, outperforming it to a minor extent after 80 trials, while in $D_6$ and $D_8$, its F1 ratio raises up to 0.93 after 100 trials, because there is an even lower portion of top-performing configurations. 

TPESampler exhibits a much 
quicker
convergence. With only 20 trials, its F1 ratio is higher than 0.95 in all datasets but the most challenging ones, namely $D_1$, $D_6$ and $D_8$. For 30 trials, its F1 ratio is lower than 0.95 only in two datasets: $D_1$ and $D_6$. In $D_1$, TPESampler outperforms grid search by $\sim$2.5\% after just 65 trials, while in $D_6$, its F1 ratio exceeds 0.95 after 70 trials. 

QMCSampler exhibits a performance similar to RandomSampler. After 20 trials, its F1 ratio is lower than 0.90 in only two datasets: $D_6$ and $D_8$. After 30 trials, its F1 ratio is lower than 0.95 in three datasets: $D_5$, $D_6$ and $D_8$. For the first two of these datasets, it matches the effectiveness of grid search after 55 trials, but its F1-score in $D_8$ is almost 9\% lower than grid search after 100 trials. In $D_1$, QMCSampler converges faster than all other samplers, with its F1 ratio exceeding 0.95 after 30 trials and 1.00 
after 80 trials. 

Finally, GPSampler is similar to TPESampler. After 20 trials, its F1 ratio exceeds 0.95 in all but the three most challenging datasets, i.e., $D_1$, $D_6$ and $D_8$. For the last two, the F1 ratio raises to 0.94 after 30 trials, with $D_1$ converging much slower, after 80 trials, eventually outperforming grid search after 100 trials. The same applies to a minor extent to $D_7$, too. Note also that GPSampler is the only approach that matches or exceeds the performance of grid search across all datasets after 100 trials. Hence, it is considered the top performing approach, albeit to a minor extent in most cases.

\begin{table}[t]
\centering
\setlength{\tabcolsep}{2.5pt}
\caption{The highest F1-score among the configurations considered by the sampling-based search algorithms in Section~\ref{sec:problem-1}.}
\vspace{-5pt}
\label{tab:global-bestf1s}
\begin{tabular}{|c|c|c|c|c|c|c|r|}
\hline
Dat. & LM & k & Clustering & Threshold & Sampler & F1 & RT (s) \\
\hline
\hline
D1 & st5 & 73 & CCC & 0.874280 & tpe & 78.43 & 0.95 \\
D2 & st5 & 10 & UMC & 0.594537 & tpe & 85.85 & 0.38 \\
D3 & sminilm & 10 & UMC & 0.429178 & random & 58.97 & 0.43 \\
D4 & st5 & 1 & UMC & 0.759879 & gps & 98.56 & 6.58 \\
D5 & st5 & 1 & CCC & 0.765179 & gps & 78.92 & 3.67 \\
D6 & sminilm & 1 & UMC & 0.555229 & gps & 60.42 & 1.30 \\
D7 & sminilm & 84 & CCC & 0.811881 & gps & 67.76 & 5.22 \\
D8 & st5 & 75 & KC & 0.920773 & tpe & 49.53 & 6.45 \\
D9 & st5 & 65 & KC & 0.830563 & tpe & 94.92 & 16.83 \\
D10 & st5 & 2 & KC & 0.269090 & tpe & 56.11 & 14.03 \\
\hline
\end{tabular}
\vspace{-16pt}
\end{table}

\textbf{RQ3.} To estimate the relative time efficiency of grid and sampling-based search, we define the \textit{runtime ratio} as $rt(sa, n)/rt(gs)$, where $rt(sa, n)$ stands for the overall run-time required by each sampler for $n$ trials, on average, across the 5 iterations (based on different seeds), and $rt(gs)$ for the overall run-time required by grid-search for the same dataset, as reported in Table \ref{tab:best-gridesearch-trials} -- $rt(sa, n)$ and $rt(gs)$ involve both the time required for generating the next parameter configuration to be tested and the run-time of the corresponding ETEER pipeline. The runtime ratio is defined in $[0,1]$, with values $<1$ indicating 
that sampling-based is faster than grid search. 

The actual values of the runtime ratio per sampler, dataset and number of trials is depicted in Figure \ref{fig:p1-samplers-performance-f1}(b). We observe that in all cases, its value is well below 0.35\%. This means sampling-based search is consistently faster than grid search by more than 285 times. Typically, the larger a dataset is, the higher is the runtime ratio, because of the more time-consuming ETEER pipelines that have to be evaluated. The relative run-time of the sampling-based algorithms is determined by their relative time complexity: the fastest approach is RandomSampler followed in close distance by QMCSampler, given their time complexity, $O(d)$ and $O(d n)$, resp.; TPESampler is consistently slower, $O(d n logn)$, with GPSampler exhibiting the highest run-times, due to its cubic time complexity.

\begin{figure}[t]
    \centering
    \includegraphics[width=0.9\linewidth]{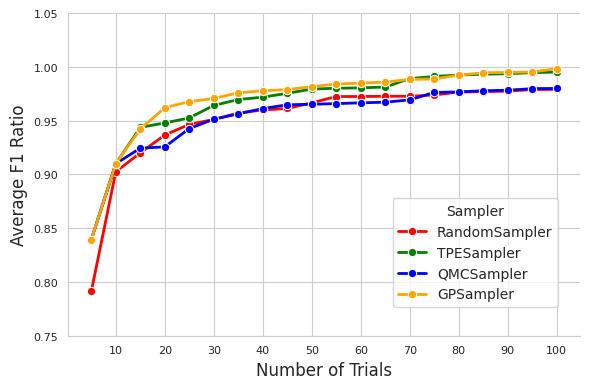}
    \vspace{-10pt}
    \caption{Average F1 score per sampler and per trial, across all the datasets.}
    \vspace{-15pt}
    \label{fig:mean-F1-samplers-performance-f1}
\end{figure}

\textbf{Conclusions.} We can conclude that all considered sampling-based search algorithms for parameter fine-tuning are capable of approximating the performance of grid search in any of the considered datasets, while being faster by at least 2 orders of magnitude.
The larger the portion of top-performing configurations in a dataset, the fewer trials are needed by these algorithms. Nevertheless, they consistently underperform grid search, albeit to an insignificant extent ($\ll0.5\%$).
Comparing their best performance in Table~\ref{tab:global-bestf1s} with the best grid search one in Table~\ref{tab:best-gridesearch-trials}, we observe that only in $D_1$ and $D_8$ the former outperforms the latter by 2\%-3\%.
As a result, sampling-based search basically 
offers a much better balance between effectiveness and time efficiency than grid search: for the same F1 score, the number of trials and the corresponding run-time is reduced by 2 or even 3 orders of magnitude. Among the sampling-based algorithms, the fastest convergence and the highest effectiveness typically correspond to GPSampler, especially for 20-30 trials, as shown in Figure \ref{fig:mean-F1-samplers-performance-f1}, which estimates the average F1 per number of trials for each sampler across all datasets in Table \ref{tab:dataset-specs}.

\section{Tackling Problem \ref{pr:pr2}}
\label{sec:problem-2}

\begin{figure*}[t]
    \centering
    \includegraphics[width=0.85\linewidth]{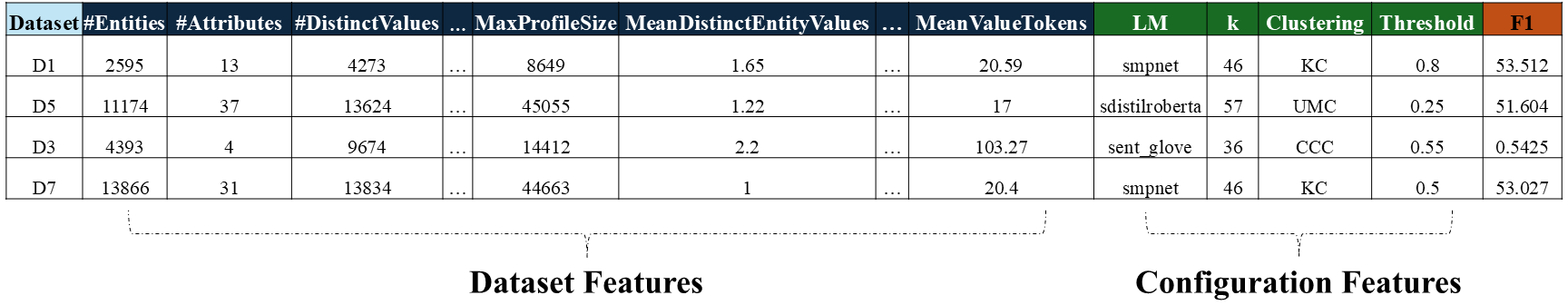}
    \vspace{-12pt}
    \caption{The feature vector of the regression model addressing Problem \ref{pr:pr2}. It combines 12 dataset features with 4 configuration parameters of the ETEER pipeline in Figure \ref{fig:eeter_pipeline}. Note that the leftmost column ("Dataset") is only added for clarification  purposes.}
    \vspace{-12pt}
    \label{fig:join_dataset_trials}
\end{figure*}

To address Problem \ref{pr:pr2}, we define a regression model that treats every combination of a dataset $D$ and a set of configuration values $V$ as an instance. For the labelled instances, the dependent variable corresponds to the F1 score of the ETEER pipeline in Figure \ref{fig:eeter_pipeline}, configured with $V$, when applied on $D$. For the unlabelled instances, the goal is to predict this F1 score. 

To put this approach into practice, we need to define the following aspects of the regression model:

\begin{itemize}[leftmargin=*]
    \item Feature engineering. This means that we need to define the feature vector that is fed to the regression model during the training and prediction phases. This vector consists of two types of dimensions: (i) \textit{dataset features}, and (ii) \textit{configuration features}. For the former, which capture the characteristics of a Record Linkage dataset, we consider features from dataset profiles defined in the literature, as explained in Section \ref{sec:datasetProfiling}. For the configuration features, the definition is straightforward: there is a separate feature for each parameter of the ETEER pipeline in Table \ref{tab:parameter-values}. The domain of each feature is the same as the respective numeric parameter, in the case of $k$ and the similarity threshold. The categorical parameters (i.e., LM and clustering algorithm) are transformed into a binary format through one-hot encoding. 
    \item Instance generation. We define three approaches for generating the labelled instances that will be used for training the regression model: (i) grid search, (ii) sampling-based search, and (iii) their combination. The resulting configuration features are combined with the dataset features extracted from the datasets D1-D10 in Table \ref{tab:dataset-specs} except for the one that should be fine-tuned. For each feature vector, we apply the respective parameter configuration to a particular dataset with known ground truth in order to estimate the corresponding target variable, i.e., the F1-score. 
    \item Learning process. We use two methods for training the regression model: (i) Random Forest and (ii) AutoML. A critical characteristic is that both methods incorporate feature selection, which is necessary due to the high dimensionality of the feature vector defined by feature engineering and the unclear contribution of each dimension. Another crucial characteristic is that they both achieve very high performance under versatile settings \cite{DBLP:journals/csur/KarmakerHSXZV22,DBLP:journals/bioinformatics/NguyenJGSALJCGM21}.
\end{itemize}

We delve into each of these three steps in the following.

\subsection{Feature Engineering}
\label{sec:datasetProfiling}

The dataset features used by our approach should adhere to the following principles: 
\begin{enumerate}[leftmargin=*]
    \item They should be \textit{generic}, applying seamlessly to any ER dataset, regardless of its format (i.e., be it structured like a CSV file or semi-structured like JSON file or RDF dump) and regardless of the corresponding flavor of ER (i.e., Record Linkage, Deduplication or Multi-source ER).
    \item They should be \textit{effective}, capturing all aspects of given dataset $D$ that might affect the performance of an ETEER pipeline on~$D$.
    \item They should be \textit{efficient}, involving low extraction cost and overhead, so that it is possible to extract predictions for numerous feature vectors from the learned model.
\end{enumerate}

In this context, we consider the following wide range of dataset features, which captures the main ones proposed in the literature. Our goal is to perform feature selection during the learning process so as to identify the top performing features for the task at hand.
\begin{enumerate}[leftmargin=*, label=F\arabic*), start=1]
    \item \#Entities \cite{DBLP:conf/sigmod/IlyasMHBA04}: the total number of entities in the dataset, i.e., $|\mathcal{E}_1| + |\mathcal{E}_2|$ in the case of Record Linkage datasets. This feature 
    affects proportionately the number of nearest neighbors returned for each query entity.
    \item \#Attributes: The total number of distinct attributes describing all given entities in $\mathcal{E}_1$ and $\mathcal{E}_2$. This is an indication of the schema heterogeneity of the given dataset(s), with higher values indicating more heterogeneous and thus noisy datasets. This is called schema complexity in \cite{DBLP:conf/cikm/PrimpeliB20}.
    \item \#DistinctValues \cite{DBLP:conf/sigmod/IlyasMHBA04}: the total number of distinct values across all attributes. Higher values suggest a larger diversity in the description of entity profiles, which indicates more challenging settings for an ETEER pipeline.
    \item \#AttributeValuePairs~\cite{DBLP:journals/tbd/EfthymiouSC20}: the total number of attribute-value pairs in all entity profiles of the given dataset(s). Higher values indicate larger profile sizes, which are probably harder to match.
    \item MeanProfileSize~\cite{DBLP:journals/tbd/EfthymiouSC20}: the average number of attribute-value pairs per entity, i.e., F4/F1. The rationale is the same as F4.
    \item MeanAttributeSize: the average number of name-value pairs per attribute, i.e., F4/F2. Higher values indicate attributes with a larger domain, thus diversifying the entity descriptions and rendering ER a more challenging task.
    \item MeanDistinctEntityValues: the average number of distinct values per entity, i.e., F3/F1. Lower values indicate profiles with insufficient or repeated information, thus hampering ER.
    \item MeanDistinctAttributeValues~\cite{DBLP:journals/pvldb/SuchanekAS11}: the average number of distinct values per attribute, i.e., F3/F2. Low values indicate datasets with non-distinctive attribute-value pairs, which are thus harder to deduplicate.
    \item MaxProfileSize: the maximum number of attribute-value pairs in any of the given entities. Higher values indicate harder entity matching settings, e.g., datasets with oversized profiles, which are probably associated with irrelevant and, thus, noisy values. 
    \item MissingInformation: The total number of missing attribute-value pairs in the given dataset(s), which is estimated as F1 $\times$ F7 - F4. Higher values indicate higher levels of noise in the given datasets, which hamper ER. This is called sparsity in~\cite{DBLP:conf/cikm/PrimpeliB20}.
    \item MeanValueTokens: The average number of tokens in all attribute values per entity. Lower values correspond to datasets with small entity descriptions, which probably lack sufficient information for ER. This is called textuality in~\cite{DBLP:conf/cikm/PrimpeliB20}.
    \item MeanValueLength: This is a variation of MeanValueTokens (or textuality \cite{DBLP:conf/cikm/PrimpeliB20}). It concatenates all attribute values per entity in a sentence, but instead of counting the words formed by tokenizing it on whitespace, it measures its length in characters. This length is averaged, across all input entities. Longer values indicate richer entity profiles, which might involve more distinguishing information, facilitating their deduplication.
\end{enumerate}

Note that all dataset features satisfy the requirements defined above, being generic, effective and efficient. They are concatenated with the four configuration features of the ETEER pipeline in Figure~\ref{fig:eeter_pipeline}, forming the 16-dimensional feature vector in Figure~\ref{fig:join_dataset_trials}.

\subsection{Instance Generation}
\label{sec:instanceGeneration}


We follow three different procedures for generating a set of labelled instances from every dataset:
\begin{enumerate}[leftmargin=*]
    \item \textit{Grid search} applies all configurations in Table \ref{tab:parameter-values} to estimate the dependent variable per instance, i.e., the respective F1-score.
    \item \textit{Sampling-based search} applies the four sampling methods in Section \ref{sec:problem-1} for a specific number of trials. F1-score is only estimated for these trials, yielding an equal number of labelled instances.
    \item \textit{All} merges the instances generated by the two approaches.
\end{enumerate}

In all cases, we exclude instances with a zero F1 as well as duplicate instances, which have the same configuration (e.g., proposed by different samplers).
Note that instance generation is a time-consuming process, due to the large number of ETEER pipelines that are evaluated. However, it constitutes an offline process that is carried out only once and, thus, does not affect the prediction time when applying the trained regression model on a new dataset that lacks a ground truth.


\subsection{Learning process}
\label{sec:learningProcess}


Having generated labelled instances from datasets with known ground truth, we train a regression model that estimates the F1-score per feature vector (i.e., per pipeline configuration and dataset without ground truth), in two different ways:
\begin{enumerate}[leftmargin=*]
    \item Random Forest (RF) \cite{DBLP:conf/icdar/Ho95}. This a simple, yet effective, approach to learn an ensemble of regression models. RF contains five hyperparameters whose tuning is essential: n\_estimators $\in \{100, ..., 1000\} \subseteq \mathbb{Z}$,  max\_depth  $ \in \{3,...,10\} \subseteq \mathbb{Z}$, min\_samples\_split $\in \{2,...,20\} \subseteq \mathbb{Z}$, min\_samples\_leaf $\in \{1,...,20\} \subseteq \mathbb{Z}$, max\_features $\in \{ sqrt, log2\}$.
    To fine-tune within these domains, we leverage Optuna as follows: First, we split the labelled instances into train and validation sets, with the latter created in a stratified manner so that it contains 10\% of the instances from each known dataset. Next, Optuna uses the validation set to find a near-optimal parameter set within 50 trials, by minimizing the mean squared error. 
    \item AutoML \cite{DBLP:journals/kbs/HeZC21}. 
    This approach evaluates a wide range of ML algorithms, such as decision trees and neural networks, without any human intervention. It also performs automatic hyperparameter optimization, considering models and configurations that might be overlooked during a manual process. For this reason, it typically yields higher accuracy than any manual procedure.
\end{enumerate}

Note that both RF and AutoML learn train regression models on subsets of the features of Section~\ref{sec:datasetProfiling}. That is, both inherently apply \textit{feature selection}. The main difference is that RF assigns the same weight to each individual model, taking the average of their predictions, whereas AutoML considers a weighted arithmetic mean. In more detail,
AutoML first explores and optimizes different regression models and then uses their predictions on the validation set 
to build an ensemble from the top N performing models \cite{DBLP:conf/icml/CaruanaNCK04} -- in our case, N goes up to 5. In fact, 
a weight is assigned to each model in proportion to its performance on the validation set, so that better-performing models receive greater weights. To make its final predictions, AutoML computes a weighted average of the predictions from the individual models.
\section{Experimental Evaluation - Problem 2}
\label{sec:expProblem2}

\begin{figure*}[t]
    \centering
    \includegraphics[width=0.9\textwidth]{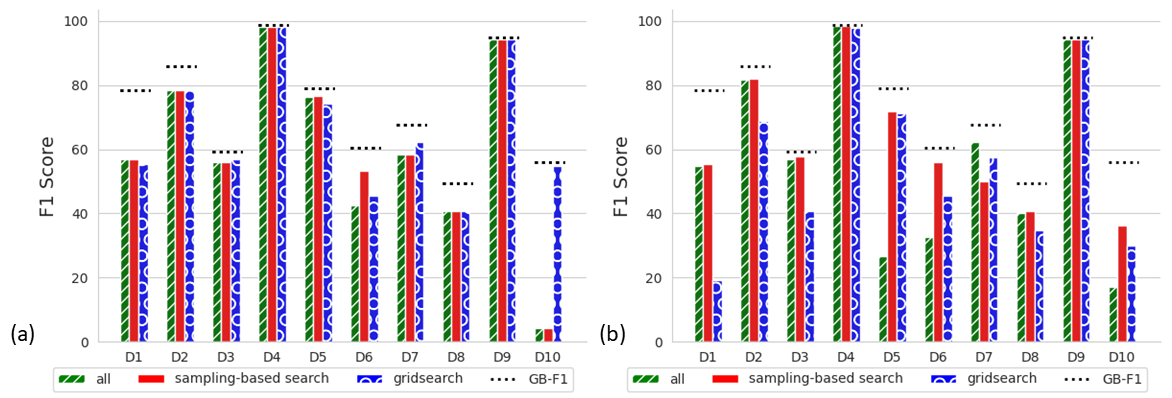}
    \vspace{-10pt}
    \caption{The F1-score of (a) the trained RF, and (b) the trained AutoML model per dataset and instance generator. GB-F1 stands for the globally best F1-score among the search algorithms in Tables \ref{tab:best-gridesearch-trials} and \ref{tab:global-bestf1s}, i.e., the performance of the best search pipeline.}
    \vspace{-14pt}
    \label{fig:rfAutoML}
\end{figure*}
\vspace{-3pt}
\subsection{Experimental Setup}\label{ssec:setup-p2}
For the implementation of Random Forest, we used scikit-learn v. 1.4.2\footnote{\url{https://scikit-learn.org}}. 
For
AutoML, we used auto-sklearn v. 1.4.2\footnote{\url{https://automl.github.io/auto-sklearn}} with the following parameters: (i) The time limit in seconds for the search of appropriate models (parameter: time\_left\_for\_this\_task) was set to 12 hours. (ii) The time for a single call in the ML model (parameter: per\_run\_time\_limit) was set to 4 hours. (iii) The memory limit in MB for the machine learning algorithm (parameter: memory\_limit) equal to  24.5 GB.
(iv) The number of jobs to run in parallel (parameter: n\_jobs) was set to 1.
All experiments were executed on a server running Ubuntu 22.04, equipped with Intel Xeon E5-4603 v2 @ 2,2GHz and 16 GM RAM.

\vspace{-3pt}

\subsection{Evaluation Results}
\label{sec:tackleProblem2}
\vspace{-3pt}
To fine-tune the workflow in Figure~\ref{fig:eeter_pipeline} without any indication of matches, we apply the following procedure, resembling the leave-one-out cross-validation approach: for each dataset $D_x$ in Table~\ref{tab:dataset-specs}, we use as training set all instances generated by grid and/or sampling-based search for all other nine datasets (i.e., all datasets among $D_1$ and $D_{10}$, except $D_x$). Using these labelled instances, we train a regression model using one of the approaches in Section \ref{sec:learningProcess}. Then, we apply the learned regression model to all instances generated by the same approach for $D_x$, estimating the respective F1-score. The instance corresponding the maximum predicted F1 provides the configuration features for the ETEER pipeline that will be eventually applied to $D_x$ in order to compute the actual F1-score.

It is important to clarify the following points: 
(i) In order to find the optimal parameters, we train a model that \emph{predicts} F1-scores per configuration and picks the best one. 
(ii) After picking the best configuration, we evaluate it by computing the \emph{actual} F1 score.
(iii) Thus, we are not actually interested in the accuracy of our F1-score predictions; we are only interested in the actual F1-scores computed by using our suggested configurations.


In this context, we address the following research questions:
\begin{enumerate}[leftmargin=*, label=RQ\arabic*), start=1]
    \item Do the three instance generation approaches in Section \ref{sec:instanceGeneration} affect the performance of the learned models?
    \item Which of the two learning processes in Section \ref{sec:learningProcess} exhibits the highest effectiveness and time efficiency?
    \item Do all features in Section \ref{sec:datasetProfiling} contribute to the effectiveness of the solutions to Problem 2?
    \item Do  our solutions to Problem 2  outperform the baseline methods w.r.t. effectiveness and time efficiency?
    \item Do our solutions to Problem 2 generalize to an unseen dataset of completely different characteristics and scale? 
\end{enumerate}
We elaborate on these research questions in the following.

\textbf{RQ1.} Figure \ref{fig:rfAutoML}(a) reports the F1-score of RF per dataset and instance generation approach. We observe that in most cases, there are no significant differences between the three instance generators. In fact, the difference between the maximum and the minimum F1 among the three instance generators is far below 1.5\% in six of the datasets, raising to just 2.6\% and 3.6\% in $D_5$ and $D_7$, respectively. This means that there are only two datasets with significant differences between the three approaches: $D_6$, where sampling-based search takes a clear lead, and $D_{10}$, where grid search is the only approach with high performance. The extremely low performance of sampling-based search and \textit{all} on $D_{10}$ leads to \textit{grid search constituting the overall best approach for RF}, as it exhibits the highest mean F1 and the lowest variance.

Much lower robustness with respect to instance generation is observed for AutoML, as shown in Figure \ref{fig:rfAutoML}(b). There is negligible deviation between the three approaches only in the two bibilographic datasets ($D_4$ and $D_9$), where the difference between the maximum and the minimum F1 is lower than 1\%. In $D_8$, this difference raises to 6\%, while in the remaining six datasets, it ranges from 12\% to 45\%. Interestingly, there is a clear winner among the three approaches: \textit{sampling-based search achieves the highest F1 in most of the datasets, while scoring the highest mean F1 and the lowest variance.} 
This means that for AutoML, the 18,000 sampler instances (9 datasets $\times$ 4 search algorithms $\times$ 100 trials $\times$ 5 seeds) convey more useful information and less noise than the 359,100 grid search instances (9 datasets $\times$ 39,900 trials). 


\begin{table*}[t]
\begin{center}  
\setlength{\tabcolsep}{3.2pt}
\caption{Performance of AutoML on Problem 2 when combined with sampling-based instances. GB stands for Gradient Boosting \cite{Friedman2001GreedyFA_GB}, KNN for K-Nearest Neighbor \cite{Fix1989DiscriminatoryA_KNN} RF for Random Forest \cite{Breiman2001RandomF_RF} and ET for Extra Trees \cite{Geurts2006ExtremelyRT_ET}. In all cases, the search/training time is 12 hours and is omitted for brevity.} 
\vspace{-10pt}
{\small
\label{tab:autosklearn-results}
\begin{tabular}{|c|l|c|c|c|c|c|c|c|}
\hline
Dataset &  \multicolumn{1}{c|}{Learned ensemble} & F1 & 
Prediction time (s)& ETEER time (s)& LM & k & Clustering & Threshold \\
\hline
\hline
D1 & 0.38ET + 0.36RF + 0.14GB + 0.1KNN + 0.02GB & 55.35 & 240 & 5.84 & st5 & 1 & UMC & 0.2659 \\
D2 & 0.82ET + 0.12GB + 0.04RF + 0.02KNN & 81.94 & 179 & 0.18 & st5 & 1 & KC & 0.1722 \\
D3 & 0.54ET + 0.16ET + 0.14RF + 0.12GB + 0.04KNN & 57.79 & 218 & 2.98 & st5 & 32 & UMC & 0.05 \\
D4 & 0.5ET + 0.38RF + 0.08GB + 0.04KNN & 98.38 & 215 & 0.83 & st5 & 1 & KC & 0.05 \\
D5 & 0.78ET + 0.1GB + 0.06RF + 0.06KNN & 71.97 & 249 & 2.77 & smpnet & 1 & KC & 0.6659 \\
D6 & 0.4ET + 0.18RF + 0.18GB + 0.12GB + 0.12KNN & 55.98 & 253 & 1.23 & sminilm & 1 & KC & 0.6307 \\
D7 & 0.5ET + 0.24RF + 0.2GB + 0.04KNN + 0.02GB & 50.14 & 227 & 2.28 & smpnet & 1 & UMC & 0.228 \\
D8 & 0.52ET + 0.2RF + 0.12KNN + 0.1GB + 0.06GB & 40.77 & 230 & 20.66 & st5 & 91 & KC & 0.0501 \\
D9 & 0.52ET + 0.28RF + 0.14GB + 0.06KNN & 94.37 & 222 & 61.16 & st5 & 100 & KC & 0.4767 \\
D10 & 0.78ET + 0.12GB + 0.1ET & 36.31 & 68 & 13.65 & smpnet & 1 & KC & 0.05 \\
\hline
\end{tabular}
}
\end{center}  
\vspace{-15pt}
\end{table*}


\begin{table}[t]
{\small
\begin{center}  
\setlength{\tabcolsep}{4pt}
\caption{Performance of Random Forest on Problem 2 when combined with grid search instances.} 
\vspace{-10pt}
\label{tab:lr-with-data-features}
\begin{tabular}{|c|c|c|c|c|c|c|c|c|}
\hline
\multirow{2}{*}{Dat.}  & \multirow{2}{*}{F1} & Train.  & Predic.  & ETEER  & \multirow{2}{*}{LM} & \multirow{2}{*}{k} & Clust & Thre \\
 & & time (s) & time (s) &  time (s) &  &  & ering & shold \\
\hline
\hline
D1 & 55.35 & 73.10 & 1.21 & 4.11 & st5 & 1 & KC & 0.20 \\
D2 & 78.46 & 10.10 & 0.18 & 5.10 & st5 & 16 & KC & 0.10 \\
D3 & 56.86 & 10.22 & 0.18 & 0.55 & st5 & 2 & KC & 0.60 \\
D4 & 98.24 & 65.26 & 1.00 & 1.70 & st5 & 2 & KC & 0.20 \\
D5 & 74.16 & 53.34 & 0.94 & 2.05 & sminilm & 1 & KC & 0.70 \\
D6 & 45.44 & 42.50 & 0.51 & 1.20 & sminilm & 1 & KC & 0.70 \\
D7 & 62.12 & 38.81 & 0.47 & 1.58 & sminilm & 1 & KC & 0.60 \\
D8 & 40.78 & 24.27 & 0.28 & 33.10 & st5 & 99 & KC & 0.25 \\
D9 & 94.35 & 42.14 & 0.54 & 60.76 & st5 & 74 & KC & 0.35 \\
D10 & 54.91 & 48.67 & 0.67 & 37.55 & st5 & 27 & KC & 0.50 \\
\hline
\end{tabular}
\end{center}  
}
\vspace{-14pt}
\end{table}

\textbf{RQ2.} We now compare the two learning processes in combination with their best instance generation approach. The performance of Random Forest with grid search instances along with the selected workflow configurations 
is reported in Table \ref{tab:lr-with-data-features}. The same information for AutoML in combination with sampling-based instances is reported in Table \ref{tab:autosklearn-results}. We observe significant variations between the selected configurations of the two learners for each dataset.

Regarding their effectiveness, the two learners exhibit practically identical F1 score in half the datasets. In fact, the difference in their F1 score is far less than 1\% in $D_1$, $D_3$, $D_4$, $D_8$ and $D_9$. AutoML takes the lead in $D_2$ and $D_6$, while Random Forest outperforms it in the three remaining datasets. With respect to average F1 score, Random Forest takes a significant lead (66.1 vs 64.3), while exhibiting lower variance. It should be stressed that \textit{AutoML's performance depends heavily on the available search/training time}, with higher budgets yielding even better results. Yet, the current limit of 12 hours is already too high when compared to the ETEER runtime, which does not exceed 36 seconds in any of the datasets.

Regarding time efficiency, Random Forest is clearly the top performer. Its training time is consistently lower than two seconds, unlike the 12 hours required by AutoML. Its prediction time is also extremely fast, consistently requiring less than 20 milliseconds, whereas AutoML typically takes a few minutes. This should be attributed to the simpler models learned by Random Forest, unlike the complicated, weighted ensemble learned by AutoML. 

For these reasons, Random Forest trained on grid search instances is the preferred approach to tackle Problem 2 and answer RQ3. Based on RF, the overall approach for training the regression model on an existing dataset, generating the testing instances to yield a workflow configuration and applying it to a specific dataset takes less than three minutes in all datasets of Table \ref{tab:lr-with-data-features}. 

\begin{figure}[t]
    \centering
    \includegraphics[width=0.75\linewidth]{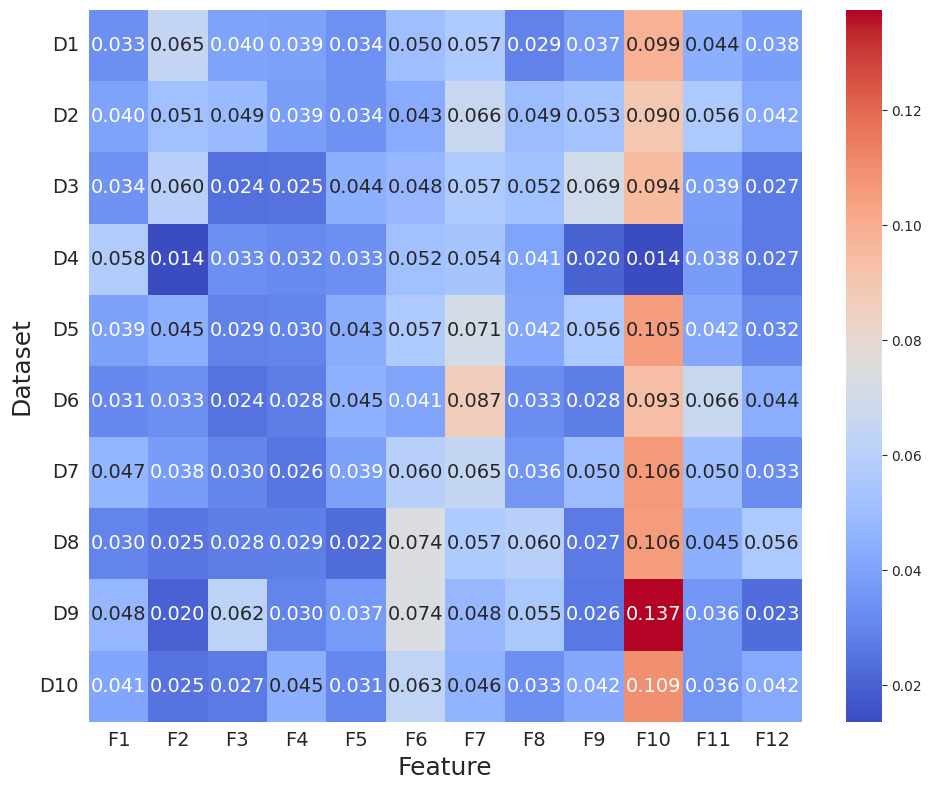}
    \vspace{-12pt}
    \caption{The gini importance of each feature per dataset for Random Forest with grid search instances.}
    \vspace{-18pt}
    \label{fig:gini}
\end{figure}

\textbf{RQ3.} A major advantage of Random Forest is the interpretability of its trained decision trees, which allows for examining the specific features that are actually useful in predicting the best workflow configuration per dataset. In fact, Gini importance provides normalized estimations of the importance of each feature in $[0,1]$, with higher values indicating more important features. 

Figure \ref{fig:gini} reports this measure for all features of Section \ref{sec:datasetProfiling} across all datasets when training Random Forest with grid search instances. \textit{All features have a non-zero importance} that fluctuates between 0.137 and 0.014. On average, the most important features are F10 (0.095), F7 (0.061) and F6 (0.056), while the least important ones are F4 0.032) and F3 (0.035). Hence, \textit{despite the seemingly small variations, some features are two or even three times more important than others}. Random Forest inherently addresses these variations without requiring a specialized feature selection approach.

\textbf{RQ4.} We now compare the top learned models, i.e.,  RF in combination with grid search instances and AutoML in combination with sampling-based search, with the following three baselines methods:

(i) The default pipeline in Section \ref{sec:tackleProblem1}
. In most datasets, this approach underperforms both RF and AutoML to a significant extent, due to its fixed configuration. The only exception is $D_2$, where the default configuration is almost the same as the best one. AutoML performs much worse than the default configuration in $D_{10}$, too, due to the high levels of noise and the low portion of top performing configurations (see Figure \ref{fig:f1_boxplot_all}), which call for much higher search times. Overall, \textit{the adaptive workflow configurations proposed by RF and AutoML typically outperform the top-performing default one}.

(ii) The best search pipeline, i.e., GB-F1 in Figure \ref{fig:rfAutoML}, which corresponds to the best performance of the ETEER pipeline in Figure~\ref{fig:eeter_pipeline} for a specific dataset among the grid and sampling-based search algorithms in Tables \ref{tab:best-gridesearch-trials} and \ref{tab:global-bestf1s}, respectively.
We observe that neither RF in combination with grid search instances nor AutoML in combination with sampling-based search instances outperform GB-F1 in any dataset. Yet, both RF and AutoML almost match GB-F1 in three datasets: $D_3$, $D_4$ and $D_9$. This is expected for the last two, the relatively clean bibliographic datasets, given the large portion of workflow configurations with very high performance in Figure \ref{fig:f1_boxplot_all}. For $D_{3}$, this shows the high effectiveness of both RF and AutoML. The worst performance of RF and AutoML with respect to this baseline corresponds to $D_1$ and $D_8$, where their F1-score is lower by $\sim$30\% and $\sim$20\%, respectively. These are the datasets with the lowest portion of duplicate entities, causing the selected workflow configurations to suffer from low precision. AutoML performs poorly in $D_{10}$, too, for reasons explained above.
All other datasets lie in the middle of these two extremes, with RF and AutoML underperforming GB-F1 by 10\% and 11\%, on average, respectively. 


(iii) ZeroER \cite{DBLP:conf/sigmod/WuCSCT20}, an established ETEER approach that involves both Filtering and Verification, while requiring no labelled instances for the dataset at hand. Its performance is reported in Table \ref{tab:zeroer-results}. Note that it did not terminate in three datasets within 2 days.
Compared to ZeroER, RF is consistently much faster: it requires far less than 3 minutes in all cases, while ZeroER requires at least 26 min in all datasets but the smallest one, where it actually finds no matches. The reason is that ZeroER cannot support missing and misplaced values, which abound in $D_1$. Apart from this dataset, RF significantly outperforms ZeroER in terms of effectiveness in three more datasets: $D_2$, $D_4$ and $D_9$. The reason is that those datasets convey long textual values, which are ideal for the pre-trained language model that lies at the core the ETEER pipeline in Figure \ref{fig:eeter_pipeline}. In contrast, $D_5$, $D_7$ and $D_{10}$ convey short textual values that usually correspond to person names. The language models struggle to find semantic similarities in these settings, unlike the string similarity measures that lie at the core of ZeroER. As a result, RF significantly underperforms ZeroER in these three datasets, but remains faster by orders of magnitude.


Compared to ZeroER, AutoML achieves higher F1-score in the same four datasets as RF, while undeperforming in the same three datasets, for the same reason (i.e., the length of attribute values). Yet, AutoML is much worse than ZeroER in $D_5$ and $D_7$, while its run-time is significantly higher than ZeroER in all datasets, but $D_9$ (and the three datasets where ZeroER runs for more than 48 hours). 

Overall, \textit{the relative effectiveness of RF and AutoML with respect to ZeroER depends on data characteristics. RF is more scalable and time efficient, while AutoML has a similar, but adjustable run-time}.

\begin{table}[t]
\setlength{\tabcolsep}{3pt}
\footnotesize
\centering
\caption{Performance over $D_{11}$ in the context of Problem 2.
}
\vspace{-8pt}
\label{tab:dbpedia-results}
\begin{tabular}{|c|c|c|c|c|c|c|c|c|}
\hline
\multirow{2}{*}{LM} & \multirow{2}{*}{K} & Cluste & Thres & Training & Prediction & F1 & ETEER  & Inst. \\
 & & ring & hold& Time (h) & Time (s) & score & Runtime (h) &  gen. \\
\hline
\hline
st5 & 10 & UMC & 0.50 & - & - & 83.22 & 14.66 & - \\
\hline
\multicolumn{9}{c}{(a) Default configuration}\\
\hline
st5 & 51 & KC & 0.15 & 0.02 & 1.00 & 84.89 & 18.04 & grid \\
st5 & 1 & KC & 0.55 & 0.00 & 0.45 & 84.83 & 13.37 & sampl. \\
st5 & 100 & KC & 0.30 & 0.03 & 0.61 & 84.98 & 15.97 & all \\
\hline
\multicolumn{9}{c}{(b) Random Forest}\\
\hline
smpnet & 85 & KC & 0.55 & 12 & 17.22 & 74.26 & 16.93 & grid. \\
st5 & 1 & KC & 0.05 & 12 & 15.54 & 84.84 & 13.40 & sampl. \\
st5 & 18 & KC & 0.05 & 12 & 13.07 & 84.84 & 16.93 &  all \\
\hline
\multicolumn{9}{c}{(c) AutoML}
\end{tabular}
\vspace{-15pt}
\end{table}


\textbf{RQ5.} In this experiment, we evaluate the generalization of the solutions to Problem 2 on $D_{11}$, a dataset with characteristics that are substantially different from all other datasets in Table \ref{tab:dataset-specs} -- unlike the limited size and schema of the other datasets, it contains millions of heterogeneous entities with user-generated content using 50,000 different attributes from two versions of DBpedia that chronologically differ by 3 years \cite{DBLP:journals/is/PapadakisMGSTGB20}.
Due to its size, every run of the ETEER pipeline in Figure \ref{fig:eeter_pipeline} typically takes $\sim$12 hours.
Hence, ZeroER and grid search are inapplicable,
while the sampling-based approaches of Section~\ref{sec:problem-1} are extremely time consuming. Therefore, we use the default configuration defined in Section \ref{sec:tackleProblem1} as baseline.

Table \ref{tab:dbpedia-results} reports the performance of this baseline
along with 
RF and AutoML in combination with all instances generated from $D_1$-$D_{10}$ by the three approaches in Section \ref{sec:instanceGeneration}. 
All tested pipelines exhibit relatively high effectiveness, with the default one matching the performance of the fine-tuned pipeline in \cite{DBLP:journals/is/PapadakisMGSTGB20}, which leverages traditional, string similarities for Filtering and Verification. This suggests that DBpedia entails many top performing configurations, like $D_4$ and $D_9$ in Figure~\ref{fig:f1_boxplot_all}. Most importantly, the configurations proposed by RF and AutoML exhibit higher F1-score by 1.6\% -- the only exception is AutoML with grid search instances, which significantly underperforms the baseline method, probably because it  requires a higher search time. This verifies the high effectiveness of the solutions to Problem 2 even in settings significantly different from the datasets generating the training instances. 

In terms of time efficiency, the prediction times of both RF and AutoML are quite low, while the run-time of the automatically configured workflows is comparable to that of the baseline.
In fact, the sampling-based instances yield lower ETEER times by 8.9\% for both RF and AutoML. The latter, though, exhibits very high training times (12 hours), yielding a much higher overall run-time.

Overall, \textit{RF and AutoML 
can automatically configure 
the ETEER pipeline even for a voluminous dataset with high levels of noise and schema heterogeneity}, with RF exhibiting very high time efficiency.

\textbf{Conclusions.} Both RF and AutoML 
are capable of fine-tuning the ETEER pipeline in Figure \ref{fig:eeter_pipeline} under versatile settings. Using the former (ideally with grid search instances), the training and prediction times are minimized, while the proposed configuration typically achieves higher effectiveness than the one proposed by AutoML, despite its time-consuming search phase. 

\vspace{-5pt}
\section{Related Work}\label{sec:related_work}


There has been limited research on automatically configuring 
ER pipelines, with most relevant works
focusing on the Entity Matching step ( i.e., on Verification), rather than end-to-end ER, as in our case. More specifically, 
a transfer-learning framework that utilizes pre-trained Entity Matching models derived from extensive knowledge bases (KBs) is proposed in \cite{AutoER_WWW}. However, this approach is domain-specific, heavily relying on relevant and well-curated KBs, which limits its applicability in real-world scenarios where such KBs are unavailable. In contrast, our ETEER methodology is domain-agnostic and holistic, also covering Filtering, and can be applied across various data types. 

Another approach to constructing a purely automatic 
Entity Matching pipeline using AutoML is presented in \cite{AutoER_EDBT}, which uses auto-sklearn to predict whether a pair of entities is a match/non-match by transforming the textual description of entities through BERT-based pre-trained language models. It differs from the tasks examined in our work in that it completely disregards Filtering.

Lastly, the 
framework in \cite{AutoER_ICDE}
uses AutoML models to predict the optimal Random Forest configuration. Again, it disregards Filtering, treating it as an orthogonal problem, while the proposed solution 
is limited to a specific type of classifier for Verification. It relies on an active learning and self-training strategy, which necessitates a human-in-the-loop algorithm when ground truth pairs are scarce. In contrast, our ETEER methodology requires no human involvement.

\begin{table}[t]
\footnotesize
\centering
\setlength{\tabcolsep}{1.4pt}
\caption{Performance of ZeroER over the datasets of Table \ref{tab:dataset-specs}.}
\vspace{-10pt}
\label{tab:zeroer-results}
\begin{tabular}{|l|c|c|c|c|c|c|c|c|c|c|}
\cline{2-11}
\multicolumn{1}{c|}{} & D1 & D2 & D3 & D4 & D5 & D6 & D7 & D8 & D9 & D10 \\
\hline
\hline
RT & 4 sec & 70 min & $\gg$48 hrs & 5.5 hrs & 34 min & $\gg$48 hrs & 67 min & $\gg$48 hrs & 24.5 hrs & 26 min \\
F1 & 0.00 & 46.46 & - & 97.00 & 92.79 & - & 88.07 & - & 64.06 & 92.62 \\
\hline
\end{tabular}
\vspace{-14pt}
\end{table}
\vspace{-5pt}
\section{Conclusion}\label{sec:conclusion}

We have defined two problems for auto-configuring end-to-end entity resolution (ETEER) pipelines: one applicable to datasets for which some known matches are available for training, and another one for a setting in which there is no ground truth of matches for the given dataset. We adapted existing techniques to solve each problem, based on a state-of-the-art ETEER pipeline.

For the first problem, we adapt established search algorithms for hyperparameter optimization to ETEER. Our thorough experimental analysis over 10 popular real-world datasets demonstrates that these algorithms match or exceed the effectiveness of grid search, while reducing the number of trials and the corresponding search time by 2 or even 3 orders of magnitude. 

For the second problem, we consider 12 generic features that can be efficiently extracted from each dataset, regardless of its format and ER setting. To learn the relations between these features, labelled instances are generated from datasets with an available ground truth and used to train a regression model using Random Forest (RF) or AutoML -- both are state-of-the-art learning algorithms that inherently perform feature selection. 
Experiments over 11 real-world datasets demonstrate the high importance of most features, with the best performance in terms of effectiveness and time efficiency corresponding to RF with grid search instances. 

In the future, we plan to generalize our approach so that it is independent of the ETEER pipeline: given a dataset, we should be able to recommend the best pipeline (not necessarily the one in Figure \ref{fig:eeter_pipeline}) along with its best hyperparameters.

\bibliographystyle{ACM-Reference-Format}
\bibliography{main}


\begin{thebibliography}{52}


\ifx \showCODEN    \undefined \def \showCODEN     #1{\unskip}     \fi
\ifx \showDOI      \undefined \def \showDOI       #1{#1}\fi
\ifx \showISBNx    \undefined \def \showISBNx     #1{\unskip}     \fi
\ifx \showISBNxiii \undefined \def \showISBNxiii  #1{\unskip}     \fi
\ifx \showISSN     \undefined \def \showISSN      #1{\unskip}     \fi
\ifx \showLCCN     \undefined \def \showLCCN      #1{\unskip}     \fi
\ifx \shownote     \undefined \def \shownote      #1{#1}          \fi
\ifx \showarticletitle \undefined \def \showarticletitle #1{#1}   \fi
\ifx \showURL      \undefined \def \showURL       {\relax}        \fi
\providecommand\bibfield[2]{#2}
\providecommand\bibinfo[2]{#2}
\providecommand\natexlab[1]{#1}
\providecommand\showeprint[2][]{arXiv:#2}

\bibitem[\protect\citeauthoryear{Akiba, Sano, Yanase, Ohta, and Koyama}{Akiba et~al\mbox{.}}{2019}]%
        {OPTUNA}
\bibfield{author}{\bibinfo{person}{Takuya Akiba}, \bibinfo{person}{Shotaro Sano}, \bibinfo{person}{Toshihiko Yanase}, \bibinfo{person}{Takeru Ohta}, {and} \bibinfo{person}{Masanori Koyama}.} \bibinfo{year}{2019}\natexlab{}.
\newblock \showarticletitle{Optuna: {A} Next-generation Hyperparameter Optimization Framework}. In \bibinfo{booktitle}{\emph{{SIGKDD}}}. \bibinfo{publisher}{{ACM}}, \bibinfo{pages}{2623--2631}.
\newblock


\bibitem[\protect\citeauthoryear{Aum{\"{u}}ller, Bernhardsson, and Faithfull}{Aum{\"{u}}ller et~al\mbox{.}}{2020}]%
        {DBLP:journals/is/AumullerBF20}
\bibfield{author}{\bibinfo{person}{Martin Aum{\"{u}}ller}, \bibinfo{person}{Erik Bernhardsson}, {and} \bibinfo{person}{Alexander~John Faithfull}.} \bibinfo{year}{2020}\natexlab{}.
\newblock \showarticletitle{ANN-Benchmarks: {A} benchmarking tool for approximate nearest neighbor algorithms}.
\newblock \bibinfo{journal}{\emph{Inf. Syst.}}  \bibinfo{volume}{87} (\bibinfo{year}{2020}).
\newblock


\bibitem[\protect\citeauthoryear{Bergstra, Bardenet, Bengio, and K\'{e}gl}{Bergstra et~al\mbox{.}}{2011}]%
        {TPE2}
\bibfield{author}{\bibinfo{person}{James Bergstra}, \bibinfo{person}{R\'{e}mi Bardenet}, \bibinfo{person}{Yoshua Bengio}, {and} \bibinfo{person}{Bal\'{a}zs K\'{e}gl}.} \bibinfo{year}{2011}\natexlab{}.
\newblock \showarticletitle{Algorithms for Hyper-Parameter Optimization}. In \bibinfo{booktitle}{\emph{Advances in Neural Information Processing Systems}}, \bibfield{editor}{\bibinfo{person}{J.~Shawe-Taylor}, \bibinfo{person}{R.~Zemel}, \bibinfo{person}{P.~Bartlett}, \bibinfo{person}{F.~Pereira}, {and} \bibinfo{person}{K.Q. Weinberger}} (Eds.), Vol.~\bibinfo{volume}{24}. \bibinfo{publisher}{Curran Associates, Inc.}
\newblock
\urldef\tempurl%
\url{https://proceedings.neurips.cc/paper_files/paper/2011/file/86e8f7ab32cfd12577bc2619bc635690-Paper.pdf}
\showURL{%
\tempurl}


\bibitem[\protect\citeauthoryear{Bergstra and Bengio}{Bergstra and Bengio}{2012}]%
        {QMCSampler}
\bibfield{author}{\bibinfo{person}{James Bergstra} {and} \bibinfo{person}{Yoshua Bengio}.} \bibinfo{year}{2012}\natexlab{}.
\newblock \showarticletitle{Random Search for Hyper-Parameter Optimization}.
\newblock \bibinfo{journal}{\emph{Journal of Machine Learning Research}} \bibinfo{volume}{13}, \bibinfo{number}{10} (\bibinfo{year}{2012}), \bibinfo{pages}{281--305}.
\newblock
\urldef\tempurl%
\url{http://jmlr.org/papers/v13/bergstra12a.html}
\showURL{%
\tempurl}


\bibitem[\protect\citeauthoryear{Bergstra, Yamins, and Cox}{Bergstra et~al\mbox{.}}{2012}]%
        {TPE3}
\bibfield{author}{\bibinfo{person}{J. Bergstra}, \bibinfo{person}{D. Yamins}, {and} \bibinfo{person}{D.~D. Cox}.} \bibinfo{year}{2012}\natexlab{}.
\newblock \bibinfo{title}{Making a Science of Model Search}.
\newblock
\newblock
\showeprint[arxiv]{1209.5111}~[cs.CV]
\urldef\tempurl%
\url{https://arxiv.org/abs/1209.5111}
\showURL{%
\tempurl}


\bibitem[\protect\citeauthoryear{Breiman}{Breiman}{2001}]%
        {Breiman2001RandomF_RF}
\bibfield{author}{\bibinfo{person}{L. Breiman}.} \bibinfo{year}{2001}\natexlab{}.
\newblock \showarticletitle{Random Forests}.
\newblock \bibinfo{journal}{\emph{Machine Learning}}  \bibinfo{volume}{45} (\bibinfo{year}{2001}), \bibinfo{pages}{5--32}.
\newblock
\urldef\tempurl%
\url{https://api.semanticscholar.org/CorpusID:89141}
\showURL{%
\tempurl}


\bibitem[\protect\citeauthoryear{Caruana, Niculescu{-}Mizil, Crew, and Ksikes}{Caruana et~al\mbox{.}}{2004}]%
        {DBLP:conf/icml/CaruanaNCK04}
\bibfield{author}{\bibinfo{person}{Rich Caruana}, \bibinfo{person}{Alexandru Niculescu{-}Mizil}, \bibinfo{person}{Geoff Crew}, {and} \bibinfo{person}{Alex Ksikes}.} \bibinfo{year}{2004}\natexlab{}.
\newblock \showarticletitle{Ensemble selection from libraries of models}. In \bibinfo{booktitle}{\emph{{ICML}}}, Vol.~\bibinfo{volume}{69}. \bibinfo{publisher}{{ACM}}.
\newblock


\bibitem[\protect\citeauthoryear{Christen}{Christen}{2012}]%
        {DBLP:books/daglib/0030287}
\bibfield{author}{\bibinfo{person}{Peter Christen}.} \bibinfo{year}{2012}\natexlab{}.
\newblock \bibinfo{booktitle}{\emph{Data Matching - Concepts and Techniques for Record Linkage, Entity Resolution, and Duplicate Detection}}.
\newblock \bibinfo{publisher}{Springer}.
\newblock


\bibitem[\protect\citeauthoryear{Christophides, Efthymiou, Palpanas, Papadakis, and Stefanidis}{Christophides et~al\mbox{.}}{2021}]%
        {DBLP:journals/csur/ChristophidesEP21}
\bibfield{author}{\bibinfo{person}{Vassilis Christophides}, \bibinfo{person}{Vasilis Efthymiou}, \bibinfo{person}{Themis Palpanas}, \bibinfo{person}{George Papadakis}, {and} \bibinfo{person}{Kostas Stefanidis}.} \bibinfo{year}{2021}\natexlab{}.
\newblock \showarticletitle{An Overview of End-to-End Entity Resolution for Big Data}.
\newblock \bibinfo{journal}{\emph{{ACM} Comput. Surv.}} \bibinfo{volume}{53}, \bibinfo{number}{6} (\bibinfo{year}{2021}), \bibinfo{pages}{127:1--127:42}.
\newblock
\urldef\tempurl%
\url{https://doi.org/10.1145/3418896}
\showDOI{\tempurl}


\bibitem[\protect\citeauthoryear{Devlin, Chang, Lee, and Toutanova}{Devlin et~al\mbox{.}}{2019}]%
        {DBLP:conf/naacl/DevlinCLT19}
\bibfield{author}{\bibinfo{person}{Jacob Devlin}, \bibinfo{person}{Ming{-}Wei Chang}, \bibinfo{person}{Kenton Lee}, {and} \bibinfo{person}{Kristina Toutanova}.} \bibinfo{year}{2019}\natexlab{}.
\newblock \showarticletitle{{BERT:} Pre-training of Deep Bidirectional Transformers for Language Understanding}. In \bibinfo{booktitle}{\emph{Proceedings of the 2019 Conference of the North American Chapter of the Association for Computational Linguistics: Human Language Technologies, {NAACL-HLT} 2019, Minneapolis, MN, USA, June 2-7, 2019, Volume 1 (Long and Short Papers)}}. \bibinfo{publisher}{Association for Computational Linguistics}, \bibinfo{pages}{4171--4186}.
\newblock


\bibitem[\protect\citeauthoryear{Douze, Guzhva, Deng, Johnson, Szilvasy, Mazar{\'{e}}, Lomeli, Hosseini, and J{\'{e}}gou}{Douze et~al\mbox{.}}{2024}]%
        {DBLP:journals/corr/abs-2401-08281}
\bibfield{author}{\bibinfo{person}{Matthijs Douze}, \bibinfo{person}{Alexandr Guzhva}, \bibinfo{person}{Chengqi Deng}, \bibinfo{person}{Jeff Johnson}, \bibinfo{person}{Gergely Szilvasy}, \bibinfo{person}{Pierre{-}Emmanuel Mazar{\'{e}}}, \bibinfo{person}{Maria Lomeli}, \bibinfo{person}{Lucas Hosseini}, {and} \bibinfo{person}{Herv{\'{e}} J{\'{e}}gou}.} \bibinfo{year}{2024}\natexlab{}.
\newblock \showarticletitle{The Faiss library}.
\newblock \bibinfo{journal}{\emph{CoRR}}  \bibinfo{volume}{abs/2401.08281} (\bibinfo{year}{2024}).
\newblock


\bibitem[\protect\citeauthoryear{Efthymiou, Stefanidis, and Christophides}{Efthymiou et~al\mbox{.}}{2020}]%
        {DBLP:journals/tbd/EfthymiouSC20}
\bibfield{author}{\bibinfo{person}{Vasilis Efthymiou}, \bibinfo{person}{Kostas Stefanidis}, {and} \bibinfo{person}{Vassilis Christophides}.} \bibinfo{year}{2020}\natexlab{}.
\newblock \showarticletitle{Benchmarking Blocking Algorithms for Web Entities}.
\newblock \bibinfo{journal}{\emph{{IEEE} Trans. Big Data}} \bibinfo{volume}{6}, \bibinfo{number}{2} (\bibinfo{year}{2020}), \bibinfo{pages}{382--395}.
\newblock
\urldef\tempurl%
\url{https://doi.org/10.1109/TBDATA.2016.2576463}
\showDOI{\tempurl}


\bibitem[\protect\citeauthoryear{Fix and Hodges}{Fix and Hodges}{1989}]%
        {Fix1989DiscriminatoryA_KNN}
\bibfield{author}{\bibinfo{person}{Evelyn Fix} {and} \bibinfo{person}{Joseph~L. Hodges}.} \bibinfo{year}{1989}\natexlab{}.
\newblock \showarticletitle{Discriminatory Analysis - Nonparametric Discrimination: Consistency Properties}.
\newblock \bibinfo{journal}{\emph{International Statistical Review}}  \bibinfo{volume}{57} (\bibinfo{year}{1989}), \bibinfo{pages}{238}.
\newblock
\urldef\tempurl%
\url{https://api.semanticscholar.org/CorpusID:120323383}
\showURL{%
\tempurl}


\bibitem[\protect\citeauthoryear{Friedman}{Friedman}{2001}]%
        {Friedman2001GreedyFA_GB}
\bibfield{author}{\bibinfo{person}{Jerome~H. Friedman}.} \bibinfo{year}{2001}\natexlab{}.
\newblock \showarticletitle{Greedy function approximation: A gradient boosting machine.}
\newblock \bibinfo{journal}{\emph{Annals of Statistics}}  \bibinfo{volume}{29} (\bibinfo{year}{2001}), \bibinfo{pages}{1189--1232}.
\newblock
\urldef\tempurl%
\url{https://api.semanticscholar.org/CorpusID:39450643}
\showURL{%
\tempurl}


\bibitem[\protect\citeauthoryear{Getoor and Machanavajjhala}{Getoor and Machanavajjhala}{2012}]%
        {DBLP:journals/pvldb/GetoorM12}
\bibfield{author}{\bibinfo{person}{Lise Getoor} {and} \bibinfo{person}{Ashwin Machanavajjhala}.} \bibinfo{year}{2012}\natexlab{}.
\newblock \showarticletitle{Entity Resolution: Theory, Practice {\&} Open Challenges}.
\newblock \bibinfo{journal}{\emph{Proc. {VLDB} Endow.}} \bibinfo{volume}{5}, \bibinfo{number}{12} (\bibinfo{year}{2012}), \bibinfo{pages}{2018--2019}.
\newblock


\bibitem[\protect\citeauthoryear{Geurts, Ernst, and Wehenkel}{Geurts et~al\mbox{.}}{2006}]%
        {Geurts2006ExtremelyRT_ET}
\bibfield{author}{\bibinfo{person}{Pierre Geurts}, \bibinfo{person}{Damien Ernst}, {and} \bibinfo{person}{Louis Wehenkel}.} \bibinfo{year}{2006}\natexlab{}.
\newblock \showarticletitle{Extremely randomized trees}.
\newblock \bibinfo{journal}{\emph{Machine Learning}}  \bibinfo{volume}{63} (\bibinfo{year}{2006}), \bibinfo{pages}{3--42}.
\newblock
\urldef\tempurl%
\url{https://api.semanticscholar.org/CorpusID:15137276}
\showURL{%
\tempurl}


\bibitem[\protect\citeauthoryear{Hassanzadeh, Chiang, Miller, and Lee}{Hassanzadeh et~al\mbox{.}}{2009}]%
        {DBLP:journals/pvldb/HassanzadehCML09}
\bibfield{author}{\bibinfo{person}{Oktie Hassanzadeh}, \bibinfo{person}{Fei Chiang}, \bibinfo{person}{Ren{\'{e}}e~J. Miller}, {and} \bibinfo{person}{Hyun~Chul Lee}.} \bibinfo{year}{2009}\natexlab{}.
\newblock \showarticletitle{Framework for Evaluating Clustering Algorithms in Duplicate Detection}.
\newblock \bibinfo{journal}{\emph{Proc. {VLDB} Endow.}} \bibinfo{volume}{2}, \bibinfo{number}{1} (\bibinfo{year}{2009}), \bibinfo{pages}{1282--1293}.
\newblock


\bibitem[\protect\citeauthoryear{He, Zhao, and Chu}{He et~al\mbox{.}}{2021}]%
        {DBLP:journals/kbs/HeZC21}
\bibfield{author}{\bibinfo{person}{Xin He}, \bibinfo{person}{Kaiyong Zhao}, {and} \bibinfo{person}{Xiaowen Chu}.} \bibinfo{year}{2021}\natexlab{}.
\newblock \showarticletitle{AutoML: {A} survey of the state-of-the-art}.
\newblock \bibinfo{journal}{\emph{Knowl. Based Syst.}}  \bibinfo{volume}{212} (\bibinfo{year}{2021}), \bibinfo{pages}{106622}.
\newblock


\bibitem[\protect\citeauthoryear{Ho}{Ho}{1995}]%
        {DBLP:conf/icdar/Ho95}
\bibfield{author}{\bibinfo{person}{Tin~Kam Ho}.} \bibinfo{year}{1995}\natexlab{}.
\newblock \showarticletitle{Random decision forests}. In \bibinfo{booktitle}{\emph{{ICDAR}}}. \bibinfo{pages}{278--282}.
\newblock


\bibitem[\protect\citeauthoryear{Ilyas, Markl, Haas, Brown, and Aboulnaga}{Ilyas et~al\mbox{.}}{2004}]%
        {DBLP:conf/sigmod/IlyasMHBA04}
\bibfield{author}{\bibinfo{person}{Ihab~F. Ilyas}, \bibinfo{person}{Volker Markl}, \bibinfo{person}{Peter~J. Haas}, \bibinfo{person}{Paul Brown}, {and} \bibinfo{person}{Ashraf Aboulnaga}.} \bibinfo{year}{2004}\natexlab{}.
\newblock \showarticletitle{{CORDS:} Automatic Discovery of Correlations and Soft Functional Dependencies}. In \bibinfo{booktitle}{\emph{{SIGMOD}}}. \bibinfo{pages}{647--658}.
\newblock


\bibitem[\protect\citeauthoryear{Kir{\'{a}}ly}{Kir{\'{a}}ly}{2013}]%
        {DBLP:journals/algorithms/Kiraly13}
\bibfield{author}{\bibinfo{person}{Zolt{\'{a}}n Kir{\'{a}}ly}.} \bibinfo{year}{2013}\natexlab{}.
\newblock \showarticletitle{Linear Time Local Approximation Algorithm for Maximum Stable Marriage}.
\newblock \bibinfo{journal}{\emph{Algorithms}} \bibinfo{volume}{6}, \bibinfo{number}{3} (\bibinfo{year}{2013}), \bibinfo{pages}{471--484}.
\newblock


\bibitem[\protect\citeauthoryear{K{\"{o}}pcke, Thor, and Rahm}{K{\"{o}}pcke et~al\mbox{.}}{2010}]%
        {DBLP:journals/pvldb/KopckeTR10}
\bibfield{author}{\bibinfo{person}{Hanna K{\"{o}}pcke}, \bibinfo{person}{Andreas Thor}, {and} \bibinfo{person}{Erhard Rahm}.} \bibinfo{year}{2010}\natexlab{}.
\newblock \showarticletitle{Evaluation of entity resolution approaches on real-world match problems}.
\newblock \bibinfo{journal}{\emph{Proc. {VLDB} Endow.}} \bibinfo{volume}{3}, \bibinfo{number}{1} (\bibinfo{year}{2010}), \bibinfo{pages}{484--493}.
\newblock


\bibitem[\protect\citeauthoryear{Lacoste{-}Julien, Palla, Davies, Kasneci, Graepel, and Ghahramani}{Lacoste{-}Julien et~al\mbox{.}}{2013}]%
        {DBLP:conf/kdd/Lacoste-JulienPDKGG13}
\bibfield{author}{\bibinfo{person}{Simon Lacoste{-}Julien}, \bibinfo{person}{Konstantina Palla}, \bibinfo{person}{Alex Davies}, \bibinfo{person}{Gjergji Kasneci}, \bibinfo{person}{Thore Graepel}, {and} \bibinfo{person}{Zoubin Ghahramani}.} \bibinfo{year}{2013}\natexlab{}.
\newblock \showarticletitle{SIGMa: simple greedy matching for aligning large knowledge bases}. In \bibinfo{booktitle}{\emph{{KDD}}}. \bibinfo{pages}{572--580}.
\newblock


\bibitem[\protect\citeauthoryear{Mikolov, Chen, Corrado, and Dean}{Mikolov et~al\mbox{.}}{2013}]%
        {DBLP:journals/corr/abs-1301-3781}
\bibfield{author}{\bibinfo{person}{Tom{\'{a}}s Mikolov}, \bibinfo{person}{Kai Chen}, \bibinfo{person}{Greg Corrado}, {and} \bibinfo{person}{Jeffrey Dean}.} \bibinfo{year}{2013}\natexlab{}.
\newblock \showarticletitle{Efficient Estimation of Word Representations in Vector Space}. In \bibinfo{booktitle}{\emph{1st International Conference on Learning Representations, {ICLR} 2013, Scottsdale, Arizona, USA, May 2-4, 2013, Workshop Track Proceedings}}.
\newblock


\bibitem[\protect\citeauthoryear{Mudgal, Li, Rekatsinas, Doan, Park, Krishnan, Deep, Arcaute, and Raghavendra}{Mudgal et~al\mbox{.}}{2018}]%
        {DBLP:conf/sigmod/MudgalLRDPKDAR18}
\bibfield{author}{\bibinfo{person}{Sidharth Mudgal}, \bibinfo{person}{Han Li}, \bibinfo{person}{Theodoros Rekatsinas}, \bibinfo{person}{AnHai Doan}, \bibinfo{person}{Youngchoon Park}, \bibinfo{person}{Ganesh Krishnan}, \bibinfo{person}{Rohit Deep}, \bibinfo{person}{Esteban Arcaute}, {and} \bibinfo{person}{Vijay Raghavendra}.} \bibinfo{year}{2018}\natexlab{}.
\newblock \showarticletitle{Deep Learning for Entity Matching: {A} Design Space Exploration}. In \bibinfo{booktitle}{\emph{{SIGMOD}}}. \bibinfo{publisher}{{ACM}}, \bibinfo{pages}{19--34}.
\newblock


\bibitem[\protect\citeauthoryear{Nguyen, J{\'{e}}z{\'{e}}quel, Gillois, Silva, Azzouz, Lambert{-}Lacroix, Juin, Campone, Gaultier, Moreau{-}Gaudry, and Antonioli}{Nguyen et~al\mbox{.}}{2021}]%
        {DBLP:journals/bioinformatics/NguyenJGSALJCGM21}
\bibfield{author}{\bibinfo{person}{Jean{-}Michel Nguyen}, \bibinfo{person}{Pascal J{\'{e}}z{\'{e}}quel}, \bibinfo{person}{Pierre Gillois}, \bibinfo{person}{Luisa Silva}, \bibinfo{person}{Faouda~Ben Azzouz}, \bibinfo{person}{Sophie Lambert{-}Lacroix}, \bibinfo{person}{Philippe~P. Juin}, \bibinfo{person}{Mario Campone}, \bibinfo{person}{Aur{\'{e}}lie Gaultier}, \bibinfo{person}{Alexandre Moreau{-}Gaudry}, {and} \bibinfo{person}{Daniel Antonioli}.} \bibinfo{year}{2021}\natexlab{}.
\newblock \showarticletitle{Random forest of perfect trees: concept, performance, applications and perspectives}.
\newblock \bibinfo{journal}{\emph{Bioinform.}} \bibinfo{volume}{37}, \bibinfo{number}{15} (\bibinfo{year}{2021}), \bibinfo{pages}{2165--2174}.
\newblock


\bibitem[\protect\citeauthoryear{Nikoletos, Papadakis, and Koubarakis}{Nikoletos et~al\mbox{.}}{2022}]%
        {Nikoletos2022pyJedAIAL}
\bibfield{author}{\bibinfo{person}{Konstantinos Nikoletos}, \bibinfo{person}{George Papadakis}, {and} \bibinfo{person}{Manolis Koubarakis}.} \bibinfo{year}{2022}\natexlab{}.
\newblock \showarticletitle{pyJedAI: a Lightsaber for Link Discovery}. In \bibinfo{booktitle}{\emph{International Workshop on the Semantic Web}}.
\newblock
\urldef\tempurl%
\url{https://api.semanticscholar.org/CorpusID:253270119}
\showURL{%
\tempurl}


\bibitem[\protect\citeauthoryear{Obraczka, Schuchart, and Rahm}{Obraczka et~al\mbox{.}}{2021}]%
        {DBLP:conf/esws/ObraczkaSR21}
\bibfield{author}{\bibinfo{person}{Daniel Obraczka}, \bibinfo{person}{Jonathan Schuchart}, {and} \bibinfo{person}{Erhard Rahm}.} \bibinfo{year}{2021}\natexlab{}.
\newblock \showarticletitle{Embedding-Assisted Entity Resolution for Knowledge Graphs}. In \bibinfo{booktitle}{\emph{{KGCW} co-located at {ESWC}}}.
\newblock


\bibitem[\protect\citeauthoryear{Paganelli, Buono, Guerra, Pevarello, and Vincini}{Paganelli et~al\mbox{.}}{2021}]%
        {AutoER_EDBT}
\bibfield{author}{\bibinfo{person}{Matteo Paganelli}, \bibinfo{person}{Francesco~Del Buono}, \bibinfo{person}{Francesco Guerra}, \bibinfo{person}{Marco Pevarello}, {and} \bibinfo{person}{Maurizio Vincini}.} \bibinfo{year}{2021}\natexlab{}.
\newblock \showarticletitle{Automated Machine Learning for Entity Matching Tasks}. In \bibinfo{booktitle}{\emph{Proceedings of the 24th International Conference on Extending Database Technology (EDBT)}} (Modena, Italy) \emph{(\bibinfo{series}{Short Papers})}. OpenProceedings.org, \bibinfo{pages}{325--330}.
\newblock
\showISBNx{978-3-89318-084-4}
\urldef\tempurl%
\url{https://doi.org/10.5441/002/edbt.2021.29}
\showDOI{\tempurl}


\bibitem[\protect\citeauthoryear{Papadakis, Efthymiou, Thanos, Hassanzadeh, and Christen}{Papadakis et~al\mbox{.}}{2023}]%
        {DBLP:journals/vldb/PapadakisETHC23}
\bibfield{author}{\bibinfo{person}{George Papadakis}, \bibinfo{person}{Vasilis Efthymiou}, \bibinfo{person}{Emmanouil Thanos}, \bibinfo{person}{Oktie Hassanzadeh}, {and} \bibinfo{person}{Peter Christen}.} \bibinfo{year}{2023}\natexlab{}.
\newblock \showarticletitle{An analysis of one-to-one matching algorithms for entity resolution}.
\newblock \bibinfo{journal}{\emph{{VLDB} J.}} \bibinfo{volume}{32}, \bibinfo{number}{6} (\bibinfo{year}{2023}), \bibinfo{pages}{1369--1400}.
\newblock
\urldef\tempurl%
\url{https://doi.org/10.1007/S00778-023-00791-3}
\showDOI{\tempurl}


\bibitem[\protect\citeauthoryear{Papadakis, Mandilaras, Gagliardelli, Simonini, Thanos, Giannakopoulos, Bergamaschi, Palpanas, and Koubarakis}{Papadakis et~al\mbox{.}}{2020}]%
        {DBLP:journals/is/PapadakisMGSTGB20}
\bibfield{author}{\bibinfo{person}{George Papadakis}, \bibinfo{person}{Georgios~M. Mandilaras}, \bibinfo{person}{Luca Gagliardelli}, \bibinfo{person}{Giovanni Simonini}, \bibinfo{person}{Emmanouil Thanos}, \bibinfo{person}{George Giannakopoulos}, \bibinfo{person}{Sonia Bergamaschi}, \bibinfo{person}{Themis Palpanas}, {and} \bibinfo{person}{Manolis Koubarakis}.} \bibinfo{year}{2020}\natexlab{}.
\newblock \showarticletitle{Three-dimensional Entity Resolution with JedAI}.
\newblock \bibinfo{journal}{\emph{Inf. Syst.}}  \bibinfo{volume}{93} (\bibinfo{year}{2020}), \bibinfo{pages}{101565}.
\newblock


\bibitem[\protect\citeauthoryear{Papadakis, Skoutas, Thanos, and Palpanas}{Papadakis et~al\mbox{.}}{2021}]%
        {DBLP:journals/csur/PapadakisSTP20}
\bibfield{author}{\bibinfo{person}{George Papadakis}, \bibinfo{person}{Dimitrios Skoutas}, \bibinfo{person}{Emmanouil Thanos}, {and} \bibinfo{person}{Themis Palpanas}.} \bibinfo{year}{2021}\natexlab{}.
\newblock \showarticletitle{Blocking and Filtering Techniques for Entity Resolution: {A} Survey}.
\newblock \bibinfo{journal}{\emph{{ACM} Comput. Surv.}} \bibinfo{volume}{53}, \bibinfo{number}{2} (\bibinfo{year}{2021}), \bibinfo{pages}{31:1--31:42}.
\newblock
\urldef\tempurl%
\url{https://doi.org/10.1145/3377455}
\showDOI{\tempurl}


\bibitem[\protect\citeauthoryear{Papadakis, Svirsky, Gal, and Palpanas}{Papadakis et~al\mbox{.}}{2016}]%
        {DBLP:journals/pvldb/0001SGP16}
\bibfield{author}{\bibinfo{person}{George Papadakis}, \bibinfo{person}{Jonathan Svirsky}, \bibinfo{person}{Avigdor Gal}, {and} \bibinfo{person}{Themis Palpanas}.} \bibinfo{year}{2016}\natexlab{}.
\newblock \showarticletitle{Comparative Analysis of Approximate Blocking Techniques for Entity Resolution}.
\newblock \bibinfo{journal}{\emph{Proc. {VLDB} Endow.}} \bibinfo{volume}{9}, \bibinfo{number}{9} (\bibinfo{year}{2016}), \bibinfo{pages}{684--695}.
\newblock


\bibitem[\protect\citeauthoryear{Pennington, Socher, and Manning}{Pennington et~al\mbox{.}}{2014}]%
        {DBLP:conf/emnlp/PenningtonSM14}
\bibfield{author}{\bibinfo{person}{Jeffrey Pennington}, \bibinfo{person}{Richard Socher}, {and} \bibinfo{person}{Christopher~D. Manning}.} \bibinfo{year}{2014}\natexlab{}.
\newblock \showarticletitle{Glove: Global Vectors for Word Representation}. In \bibinfo{booktitle}{\emph{Proceedings of the 2014 Conference on Empirical Methods in Natural Language Processing, {EMNLP} 2014, October 25-29, 2014, Doha, Qatar, {A} meeting of SIGDAT, a Special Interest Group of the {ACL}}}. \bibinfo{pages}{1532--1543}.
\newblock


\bibitem[\protect\citeauthoryear{Primpeli and Bizer}{Primpeli and Bizer}{2020}]%
        {DBLP:conf/cikm/PrimpeliB20}
\bibfield{author}{\bibinfo{person}{Anna Primpeli} {and} \bibinfo{person}{Christian Bizer}.} \bibinfo{year}{2020}\natexlab{}.
\newblock \showarticletitle{Profiling Entity Matching Benchmark Tasks}. In \bibinfo{booktitle}{\emph{{CIKM}}}. \bibinfo{pages}{3101--3108}.
\newblock


\bibitem[\protect\citeauthoryear{Reimers and Gurevych}{Reimers and Gurevych}{2019}]%
        {DBLP:conf/emnlp/ReimersG19}
\bibfield{author}{\bibinfo{person}{Nils Reimers} {and} \bibinfo{person}{Iryna Gurevych}.} \bibinfo{year}{2019}\natexlab{}.
\newblock \showarticletitle{Sentence-BERT: Sentence Embeddings using Siamese BERT-Networks}. In \bibinfo{booktitle}{\emph{{EMNLP-IJCNLP}}}. \bibinfo{pages}{3980--3990}.
\newblock


\bibitem[\protect\citeauthoryear{Saeedi, Nentwig, Peukert, and Rahm}{Saeedi et~al\mbox{.}}{2018a}]%
        {DBLP:journals/csimq/SaeediNPR18}
\bibfield{author}{\bibinfo{person}{Alieh Saeedi}, \bibinfo{person}{Markus Nentwig}, \bibinfo{person}{Eric Peukert}, {and} \bibinfo{person}{Erhard Rahm}.} \bibinfo{year}{2018}\natexlab{a}.
\newblock \showarticletitle{Scalable Matching and Clustering of Entities with {FAMER}}.
\newblock \bibinfo{journal}{\emph{Complex Syst. Informatics Model. Q.}}  \bibinfo{volume}{16} (\bibinfo{year}{2018}), \bibinfo{pages}{61--83}.
\newblock


\bibitem[\protect\citeauthoryear{Saeedi, Peukert, and Rahm}{Saeedi et~al\mbox{.}}{2017}]%
        {DBLP:conf/adbis/SaeediPR17}
\bibfield{author}{\bibinfo{person}{Alieh Saeedi}, \bibinfo{person}{Eric Peukert}, {and} \bibinfo{person}{Erhard Rahm}.} \bibinfo{year}{2017}\natexlab{}.
\newblock \showarticletitle{Comparative Evaluation of Distributed Clustering Schemes for Multi-source Entity Resolution}. In \bibinfo{booktitle}{\emph{{ADBIS}}}, Vol.~\bibinfo{volume}{10509}. \bibinfo{pages}{278--293}.
\newblock


\bibitem[\protect\citeauthoryear{Saeedi, Peukert, and Rahm}{Saeedi et~al\mbox{.}}{2018b}]%
        {DBLP:conf/esws/SaeediPR18}
\bibfield{author}{\bibinfo{person}{Alieh Saeedi}, \bibinfo{person}{Eric Peukert}, {and} \bibinfo{person}{Erhard Rahm}.} \bibinfo{year}{2018}\natexlab{b}.
\newblock \showarticletitle{Using Link Features for Entity Clustering in Knowledge Graphs}. In \bibinfo{booktitle}{\emph{{ESWC}}}. \bibinfo{pages}{576--592}.
\newblock


\bibitem[\protect\citeauthoryear{Saeedi, Peukert, and Rahm}{Saeedi et~al\mbox{.}}{2020}]%
        {DBLP:conf/esws/SaeediPR20}
\bibfield{author}{\bibinfo{person}{Alieh Saeedi}, \bibinfo{person}{Eric Peukert}, {and} \bibinfo{person}{Erhard Rahm}.} \bibinfo{year}{2020}\natexlab{}.
\newblock \showarticletitle{Incremental Multi-source Entity Resolution for Knowledge Graph Completion}. In \bibinfo{booktitle}{\emph{{ESWC}}}, Vol.~\bibinfo{volume}{12123}. \bibinfo{pages}{393--408}.
\newblock


\bibitem[\protect\citeauthoryear{Santu, Hassan, Smith, Xu, Zhai, and Veeramachaneni}{Santu et~al\mbox{.}}{2022}]%
        {DBLP:journals/csur/KarmakerHSXZV22}
\bibfield{author}{\bibinfo{person}{Shubhra Kanti~Karmaker Santu}, \bibinfo{person}{Md.~Mahadi Hassan}, \bibinfo{person}{Micah~J. Smith}, \bibinfo{person}{Lei Xu}, \bibinfo{person}{Chengxiang Zhai}, {and} \bibinfo{person}{Kalyan Veeramachaneni}.} \bibinfo{year}{2022}\natexlab{}.
\newblock \showarticletitle{AutoML to Date and Beyond: Challenges and Opportunities}.
\newblock \bibinfo{journal}{\emph{{ACM} Comput. Surv.}} \bibinfo{volume}{54}, \bibinfo{number}{8} (\bibinfo{year}{2022}), \bibinfo{pages}{175:1--175:36}.
\newblock


\bibitem[\protect\citeauthoryear{Snoek, Larochelle, and Adams}{Snoek et~al\mbox{.}}{2012}]%
        {NIPS2012_GPSampler}
\bibfield{author}{\bibinfo{person}{Jasper Snoek}, \bibinfo{person}{Hugo Larochelle}, {and} \bibinfo{person}{Ryan~P Adams}.} \bibinfo{year}{2012}\natexlab{}.
\newblock \showarticletitle{Practical Bayesian Optimization of Machine Learning Algorithms}. In \bibinfo{booktitle}{\emph{Advances in Neural Information Processing Systems}}, \bibfield{editor}{\bibinfo{person}{F.~Pereira}, \bibinfo{person}{C.J. Burges}, \bibinfo{person}{L.~Bottou}, {and} \bibinfo{person}{K.Q. Weinberger}} (Eds.), Vol.~\bibinfo{volume}{25}. \bibinfo{publisher}{Curran Associates, Inc.}
\newblock
\urldef\tempurl%
\url{https://proceedings.neurips.cc/paper_files/paper/2012/file/05311655a15b75fab86956663e1819cd-Paper.pdf}
\showURL{%
\tempurl}


\bibitem[\protect\citeauthoryear{Suchanek, Abiteboul, and Senellart}{Suchanek et~al\mbox{.}}{2011}]%
        {DBLP:journals/pvldb/SuchanekAS11}
\bibfield{author}{\bibinfo{person}{Fabian~M. Suchanek}, \bibinfo{person}{Serge Abiteboul}, {and} \bibinfo{person}{Pierre Senellart}.} \bibinfo{year}{2011}\natexlab{}.
\newblock \showarticletitle{{PARIS:} Probabilistic Alignment of Relations, Instances, and Schema}.
\newblock \bibinfo{journal}{\emph{Proc. {VLDB} Endow.}} \bibinfo{volume}{5}, \bibinfo{number}{3} (\bibinfo{year}{2011}), \bibinfo{pages}{157--168}.
\newblock


\bibitem[\protect\citeauthoryear{Sun, Zhang, Hu, Wang, Chen, Akrami, and Li}{Sun et~al\mbox{.}}{2020}]%
        {DBLP:journals/pvldb/SunZHWCAL20}
\bibfield{author}{\bibinfo{person}{Zequn Sun}, \bibinfo{person}{Qingheng Zhang}, \bibinfo{person}{Wei Hu}, \bibinfo{person}{Chengming Wang}, \bibinfo{person}{Muhao Chen}, \bibinfo{person}{Farahnaz Akrami}, {and} \bibinfo{person}{Chengkai Li}.} \bibinfo{year}{2020}\natexlab{}.
\newblock \showarticletitle{A Benchmarking Study of Embedding-based Entity Alignment for Knowledge Graphs}.
\newblock \bibinfo{journal}{\emph{Proc. {VLDB} Endow.}} \bibinfo{volume}{13}, \bibinfo{number}{11} (\bibinfo{year}{2020}), \bibinfo{pages}{2326--2340}.
\newblock


\bibitem[\protect\citeauthoryear{Thirumuruganathan, Li, Tang, Ouzzani, Govind, Paulsen, Fung, and Doan}{Thirumuruganathan et~al\mbox{.}}{2021}]%
        {DBLP:journals/pvldb/Thirumuruganathan21}
\bibfield{author}{\bibinfo{person}{Saravanan Thirumuruganathan}, \bibinfo{person}{Han Li}, \bibinfo{person}{Nan Tang}, \bibinfo{person}{Mourad Ouzzani}, \bibinfo{person}{Yash Govind}, \bibinfo{person}{Derek Paulsen}, \bibinfo{person}{Glenn Fung}, {and} \bibinfo{person}{AnHai Doan}.} \bibinfo{year}{2021}\natexlab{}.
\newblock \showarticletitle{Deep Learning for Blocking in Entity Matching: {A} Design Space Exploration}.
\newblock \bibinfo{journal}{\emph{Proc. {VLDB} Endow.}} \bibinfo{volume}{14}, \bibinfo{number}{11} (\bibinfo{year}{2021}), \bibinfo{pages}{2459--2472}.
\newblock


\bibitem[\protect\citeauthoryear{Wang, Zheng, Wang, and Pei}{Wang et~al\mbox{.}}{2021}]%
        {AutoER_ICDE}
\bibfield{author}{\bibinfo{person}{Pei Wang}, \bibinfo{person}{Weiling Zheng}, \bibinfo{person}{Jiannan Wang}, {and} \bibinfo{person}{Jian Pei}.} \bibinfo{year}{2021}\natexlab{}.
\newblock \showarticletitle{Automating Entity Matching Model Development}. In \bibinfo{booktitle}{\emph{2021 IEEE 37th International Conference on Data Engineering (ICDE)}}. \bibinfo{pages}{1296--1307}.
\newblock
\urldef\tempurl%
\url{https://doi.org/10.1109/ICDE51399.2021.00116}
\showDOI{\tempurl}


\bibitem[\protect\citeauthoryear{Watanabe}{Watanabe}{2023}]%
        {TPE1}
\bibfield{author}{\bibinfo{person}{Shuhei Watanabe}.} \bibinfo{year}{2023}\natexlab{}.
\newblock \bibinfo{title}{Tree-Structured Parzen Estimator: Understanding Its Algorithm Components and Their Roles for Better Empirical Performance}.
\newblock
\newblock
\showeprint[arxiv]{2304.11127}~[cs.LG]
\urldef\tempurl%
\url{https://arxiv.org/abs/2304.11127}
\showURL{%
\tempurl}


\bibitem[\protect\citeauthoryear{Wilson, Hutter, and Deisenroth}{Wilson et~al\mbox{.}}{2018}]%
        {DBLP:conf/nips/WilsonHD18}
\bibfield{author}{\bibinfo{person}{James~T. Wilson}, \bibinfo{person}{Frank Hutter}, {and} \bibinfo{person}{Marc~Peter Deisenroth}.} \bibinfo{year}{2018}\natexlab{}.
\newblock \showarticletitle{Maximizing acquisition functions for Bayesian optimization}. In \bibinfo{booktitle}{\emph{NeurIPS 2018}}. \bibinfo{pages}{9906--9917}.
\newblock


\bibitem[\protect\citeauthoryear{Wu, Chaba, Sawlani, Chu, and Thirumuruganathan}{Wu et~al\mbox{.}}{2020}]%
        {DBLP:conf/sigmod/WuCSCT20}
\bibfield{author}{\bibinfo{person}{Renzhi Wu}, \bibinfo{person}{Sanya Chaba}, \bibinfo{person}{Saurabh Sawlani}, \bibinfo{person}{Xu Chu}, {and} \bibinfo{person}{Saravanan Thirumuruganathan}.} \bibinfo{year}{2020}\natexlab{}.
\newblock \showarticletitle{ZeroER: Entity Resolution using Zero Labeled Examples}. In \bibinfo{booktitle}{\emph{{SIGMOD}}}. \bibinfo{pages}{1149--1164}.
\newblock
\urldef\tempurl%
\url{https://doi.org/10.1145/3318464.3389743}
\showDOI{\tempurl}


\bibitem[\protect\citeauthoryear{Yang and Shami}{Yang and Shami}{2020}]%
        {DBLP:journals/ijon/YangS20}
\bibfield{author}{\bibinfo{person}{Li Yang} {and} \bibinfo{person}{Abdallah Shami}.} \bibinfo{year}{2020}\natexlab{}.
\newblock \showarticletitle{On hyperparameter optimization of machine learning algorithms: Theory and practice}.
\newblock \bibinfo{journal}{\emph{Neurocomputing}}  \bibinfo{volume}{415} (\bibinfo{year}{2020}), \bibinfo{pages}{295--316}.
\newblock


\bibitem[\protect\citeauthoryear{Zeakis, Papadakis, Skoutas, and Koubarakis}{Zeakis et~al\mbox{.}}{2023}]%
        {DBLP:journals/pvldb/ZeakisPSK23}
\bibfield{author}{\bibinfo{person}{Alexandros Zeakis}, \bibinfo{person}{George Papadakis}, \bibinfo{person}{Dimitrios Skoutas}, {and} \bibinfo{person}{Manolis Koubarakis}.} \bibinfo{year}{2023}\natexlab{}.
\newblock \showarticletitle{Pre-trained Embeddings for Entity Resolution: An Experimental Analysis}.
\newblock \bibinfo{journal}{\emph{Proc. {VLDB} Endow.}} \bibinfo{volume}{16}, \bibinfo{number}{9} (\bibinfo{year}{2023}), \bibinfo{pages}{2225--2238}.
\newblock


\bibitem[\protect\citeauthoryear{Zhao and He}{Zhao and He}{2019}]%
        {AutoER_WWW}
\bibfield{author}{\bibinfo{person}{Chen Zhao} {and} \bibinfo{person}{Yeye He}.} \bibinfo{year}{2019}\natexlab{}.
\newblock \showarticletitle{Auto-EM: End-to-end Fuzzy Entity-Matching using Pre-trained Deep Models and Transfer Learning}. In \bibinfo{booktitle}{\emph{The World Wide Web Conference}} (San Francisco, CA, USA) \emph{(\bibinfo{series}{WWW '19})}. \bibinfo{publisher}{Association for Computing Machinery}, \bibinfo{address}{New York, NY, USA}, \bibinfo{pages}{2413–2424}.
\newblock
\showISBNx{9781450366748}
\urldef\tempurl%
\url{https://doi.org/10.1145/3308558.3313578}
\showDOI{\tempurl}


\end{thebibliography}


\end{document}